\documentclass[twocolumn,aps,prb,amsmath,amssymb]{revtex4-1}
\usepackage{subfigure}
\usepackage{graphicx}% Include figure files
\usepackage{dcolumn}% Align table columns on decimal point
\usepackage{bm}% bold math
\usepackage{dsfont}
\usepackage{tikz}
\usepackage{slashed}
\newcommand{\cev}[1]{\reflectbox{\ensuremath{\vec{\reflectbox{\ensuremath{#1}}}}}}

\def \Fst{$1^\textrm{st}$ }
\def \Snd{$2^\textrm{nd}$ }

\begin{document}

%\preprint{APS/123-QED}

\title{High-order topological insulators from high-dimensional Chern insulators}% Force line breaks with \\

\author{Ioannis Petrides}
\affiliation{Institute for Theoretical Physics, ETH Zurich, 8093 Z{\"u}rich, Switzerland}
\author{Oded Zilberberg}
\affiliation{Institute for Theoretical Physics, ETH Zurich, 8093 Z{\"u}rich, Switzerland}

\date{\today}% It is always \today, today,

\begin{abstract}
	Topological insulators are a novel state of matter that share a common feature: their spectral bands are associated with a nonlocal integer-valued index, commonly manifesting through quantized bulk phenomena and robust boundary effects. 
	In this work, we demonstrate using dimensional reduction that high-order topological insulators are descendants from a chiral semimetal in higher dimensions. 
	Specifically, we analyze the descendants of an ancestor four-dimensional Chern insulator in the limit where it becomes chiral and show their relation to two-dimensional second-order topological insulators.	
	Correspondingly, the quantization of the charge accumulation at the corners of the 2D descendants is obtained and related to the topological indices -- the \Fst and \Snd Chern numbers -- of the ancestor model. 
	Our approach provides a connection between the boundary states of high-order topological insulators and topological pumps -- the latter being dynamical realizations of high-dimensional Chern insulators.
\end{abstract}
\maketitle

%\tableofcontents

Over the past decades, the unique properties of topological insulators (TIs) led to many theoretical and experimental advances~\cite{RMP_TI, RMP_TI2, ozawareview}.
%In this paradigm, energy bands are characterized by a nonlocal quantity, the topological index, manifesting through robust bulk and boundary effects. 
TIs have energy bands that are characterized by a nonlocal quantity, a topological index, which manifests through robust bulk and boundary effects.
The quantization of the topological index usually relies on the presence of local symmetries~\cite{kitaev2009periodic,ryu2010topological, altland1997nonstandard}, symmorphic or nonsymmorphic crystalline symmetries~\cite{fu2011topological,kremer2018non}, or even quasiperiodic order~\cite{kraus2012topological,kraus2016quasiperiodicity,bellissard2000hull}.
The resulting TIs are extensively studied and classified according to the presence or absence of such symmetries~\cite{kitaev2009periodic,ryu2010topological, altland1997nonstandard,shiozaki2017k,alexandradinata2016topological}.

A relationship between the topological indices in different symmetry classes and dimensions is obtained using a plethora of methods, such as K-theory~\cite{bellissard1992gap, kitaev2009periodic, prodan2015virtual}, non-linear sigma model analysis~\cite{chiu2016classification, ryu2010topological, altland1997nonstandard}, and dimensional reduction~\cite{qi2011topological}. 
Specifically, the latter implies that a Chern insulator in $d$ dimensions is related to a family of models in $d-m$ dimensions, which we dub "descendant pump family".
Archetypical examples are the (2D$\to$1D)-reduction of the 2D quantum Hall effect (QHE) to Thouless's one-dimensional topological pump~\cite{thouless1982quantized, thouless1983quantization,kraus2012ye,verbin2015topological,lohse2016thouless}, and the (4D$\to $2D)-reduction of the 4D QHE to two-dimensional topological pumps~\cite{kraus2013four,lohse2018exploring,zilberberg2018photonic}. 
Similarly, the (4D$\to $3D)-dimensional reduction procedure allows for the derivation of a $\mathds{Z}_2$ index for 3D TIs as descendants from a 4D time-reversal invariant insulator~\cite{qi2008topological}.

%%%%%%%%%%%%%%%%%%%%%%%%%%%%%%%%%%%%%%%%%%%%%%%%%%%%%%%%%%%%
\begin{figure}[t!]
	\includegraphics[width=1\linewidth]{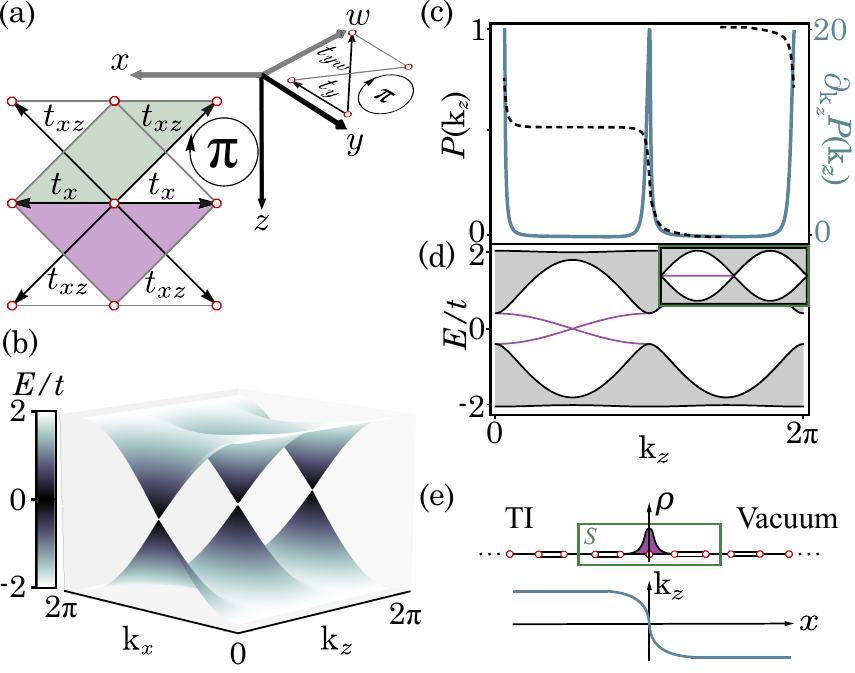} \label{fig1} 
	\caption{Ancestor model and (2D$\to$1D)-dimensional reduction. (a) The 4D hypercube model [cf.~Eq.~\eqref{HamProd}] made out of 2D Creutz lattices in the $xz$- and $yw$-planes with a $\pi$ flux threading their triangular (and parallelogram) plaquettes [cf.~Eq.~\eqref{offD}]. (b) The energy spectrum of $\hat{H} _{xz}$ [cf.~Eq.~\eqref{offD}] with periodic boundary conditions, showing two Dirac cones. 
		(c) The 1D bulk dipole $P_x (\text{k}_z)$ (dashed black line), associated with the descendant model $\hat{h}_{x}(\text{k}_z)$, and its derivative $\partial_{\text{k}_z} P_x (\text{k}_z)$ (solid blue line) as a function of $\text{k}_z$. The total area under $\partial_{\text{k}_z} P_x (\text{k}_z)$ is equal to the \Fst Chern number $c_1$ of the ancestor Creutz model [cf., Eq.~\eqref{eq:Creutz}].
		(d) The spectrum associated to $\hat{h}_{x} (\text{k}_z)$ plotted with open boundary conditions in the $x$-direction. 
		For a given $\text{k}_z$, the spectrum is projected onto the energy axis. Bulk states are shown in gray and edge states in purple. As a function of $\text{k}_z$, edge states cross the gap in correspondence with the \Fst Chern number $c_1=1$. 
		For the simulations in (c) and (d) we used $t_x=t$, $t_{xz} /t_x=0.45$ [cf.~Eq.~\eqref{eq:Creutz}] and $t_z /t_x=0.03$ to open a gap~\cite{hahaha}. 
		The inset shows the spectrum for $t_z /t_x=0$.
		(e) The area $S$ enclosing the interface between a nontrivial 1D TI and the vacuum (top) and the domain wall configuration in the $\text{k}_z$-parameter space that describes it (bottom). In addition, the charge density $\rho$ of the many-body ground state at half-filling is sketched. }
\end{figure}
%%%%%%%%%%%%%%%%%%%%%%%%%%%%%%%%%%%%%%%%%%%%%%%%%%%%%%%%%%%%

Recent research into the boundary physics of 2D materials led to the prediction and observation of zero-dimensional (0D) states, i.e., states localized in both dimensions~\cite{lin2017topological,hashimoto2017edge,langbehn2017reflection,benalcazar2017quantized,trifunovic2018higher,geier2018second,schindler2018higher,zilberberg2018photonic,wang2018higher,ezawa2018minimal,serra2018observation}.   
These states were shown to fall in a new class of TIs, dubbed ``high-order TIs", where a $d$-dimensional insulator has nontrivial boundary phenomena manifesting at its $d-m$ boundary. 
For example, a 3D second-order ($m=2$) TI has a gapped bulk spectrum and gapped 2D surfaces, but exhibits gapless topological 1D edge states~\cite{schindler2018higher}. 
The appearance of such states is understood using the modern theory of polarization extended to high multipole moments, where charge quantisation is imposed by the underlying symmetries of the system~\cite{benalcazar2017quantized}.
%A theory extending the electric dipole moment to higher multipole moments was put forward, explaining the appearance of such states through the underlying symmetries of the system.

In this paper, we use dimensional reduction to demonstrate a connection between high-order TIs and descendant pump families from high-dimensional Chern insulators. 
Specifically, we show that 2D second-order TIs are the 2D descendants of a 4D chiral semimetal.
We do so by first defining an ancestor 4D Chern insulator with well-defined \Fst- and \Snd-Chern numbers and then applying (4D$\to $2D)-dimensional reduction to obtain the descendant 2D pump family~\cite{kraus2013four,lohse2018exploring,zilberberg2018photonic}. 
In the limit where the 4D Chern insulator becomes chiral, we find that the pump family is divided into regions in parameter space separated by (bulk- or edge-) gap closures. 
These regions are distinguished by the appearance of mid-gap 0D states, localized at the corners. 
We calculate the charge accumulation at the corners of the 2D descendants using a continuoum theory and derive its quantization by connecting it to the \Snd Chern flux of the ancestor Hamiltonian. 
Using this revealed connection, we generate various 2D second-order TIs solely via flux insertions through different planes of the 4D ancestor model. 
Our results are readily generalized to any dimension, including the relation of 3D corner states to the 6D QHE and its 3$^{\text{rd}}$ Chern number~\cite{hashimoto2017edge,petrides2018six}.

We consider a tight-binding model describing spinless charged particles moving on a 4D hyper-cubic lattice in the presence of a magnetic field [see Fig.~1(a)] 
\begin{align}
\hat{H}^{4D} = \sum_{\mathbf{m}}\left[\hat{H}_{xz}(\mathbf{m})+\hat{H}_{yw}(\mathbf{m}) + \Delta\hat{H}^{b} _{xy}(\mathbf{m}) \right]\,,
\label{HamProd}
\end{align}
where $\mathbf{m}=(m_x,m_y,m_z,m_w)$ is a 4D lattice vector. The Hamiltonian density $\hat{H}_{\mu\nu}(\mathbf{m})$ describes a 2D Creutz lattice~\cite{hahaha,creutz1999m,kraus2012topological} in the $\mu\nu$-plane threaded by a magnetic field with $\pi$ flux quanta per triangular (and parallelogram) $\mu\nu$-plaquette [cf.~Fig.~\ref{fig1}(a) and (b) for the 2D Creutz lattice and its spectrum]. In the Landau gauge, the 2D Creutz models can be written as $\hat{H}_{\mu\nu}(\mathbf{m})=\hat{T}_{\mu\nu}+\hat{T}^{\dagger}_{\mu\nu} $ with
\begin{align}
\hat{T}_{\mu\nu} =  &t_{\mu\nu} \left(e^{-i\pi m_\mu }c^\dagger_{\mathbf{m}+\mathbf{e}_\mu+\mathbf{e}_\nu }c^{}_{\mathbf{m} }+e^{i\pi m_\mu }c^\dagger_{\mathbf{m}+\mathbf{e}_\mu-\mathbf{e}_\nu }c^{}_{\mathbf{m} }\right)\nonumber\\
&+t_\mu  c^\dagger_{\mathbf{m} + \mathbf{e}_\mu }c^{}_{\mathbf{m} }  \,,
\label{offD}
\end{align}
where $\mathbf{e}_\mu$ is a lattice unit vector in direction $\mu$, $t_\mu$ is the amplitude for nearest-neighbor hopping in the $\mu$-direction, $t_{\mu\nu}$ is the amplitude for next-nearest neighbor hopping, and the threaded flux is incorporated using Peierls' substitution~\cite{Peierls:1933ZPhys}. Note that for any $\frac{2\pi}{q}$-flux threading the 2D Creutz lattice, with $q$ an even integer, the low-energy theory corresponds to decoupled Dirac cones~\cite{thouless1982quantized}. 
This is crucial in defining regions in the BZ that are separated by gap closures.
The third term in Eq.~\eqref{HamProd}, $\Delta\hat{H}^{b} _{xy}(\mathbf{m})$, denotes the threading of $b=0$ or $\pi$ magnetic flux through each square plaquette in the $xy$-plane and can be written as
\begin{align}
\label{fluxXY}
\Delta\hat{H}^{b} _{xy} ({\mathbf{m}})=(e^{im_x b}-1) \hat{T}_{yw} + h.c.\,.
\end{align}
%Here too, $b$ could be taken as a $\frac{2\pi}{q}$-flux with $q$ an even integer. 
Finally, a staggered potential $\hat{V} ({\mathbf{m}})= (-1)^{m_x + m_y} \mu_0  c^\dagger_{\mathbf{m}  }c^{}_{\mathbf{m} }$ with a constant mass term $\mu_0$ (or similarly when $t_z\ne0\ne t_w$) gaps the spectrum and the resulting 4D model is a Chern insulator with well defined \Fst and \Snd Chern numbers. 

The chosen gauge in Eq.~\eqref{offD} leaves the Hamiltonian $\hat{H}^{4D}$ invariant under translations in the $z$- and $w$-direction. 
We can, therefore, write Eq.~\eqref{HamProd} in terms of the lattice quasimomenta $\tilde{\mathbf{k}}=(\text{k}_z,\text{k}_w)$,
\begin{align}
\hat{H}^{4D} = \sum_{\tilde{\mathbf{m}},\tilde{\mathbf{k}}}\left[\hat{H}_{xz}(\tilde{\mathbf{m}},\tilde{\mathbf{k}} )+\hat{H}_{yw}(\tilde{\mathbf{m}},\tilde{\mathbf{k}}) + \Delta\hat{H}^{b} _{xy} (\tilde{\mathbf{m}},\tilde{\mathbf{k}})\right]\,,
\label{HamProdK}
\end{align}
where $\tilde{\mathbf{m}}=(m_x,m_y)$,  
\begin{align}
\hat{H}_{\mu\nu}(\tilde{\mathbf{m}},\tilde{\mathbf{k}}  )=& J^-_{\mu\nu} c^\dagger_{\tilde{\mathbf{m}}+\mathbf{e}_\mu,\tilde{\mathbf{k}}}c^{}_{\tilde{\mathbf{m}},\tilde{\mathbf{k}} } +J^+_{\mu\nu}  c^\dagger_{\tilde{\mathbf{m}}-\mathbf{e}_\mu,\tilde{\mathbf{k}}  } c^{}_{\tilde{\mathbf{m}},\tilde{\mathbf{k}} } \,,\label{eq:Creutz}\\
\Delta\hat{H}^{b} _{xy} (\tilde{\mathbf{m}},\tilde{\mathbf{k}})=&(e^{im_x b}-1) J^-_{yw} c^\dagger_{\tilde{\mathbf{m}}+\mathbf{e}_y,\tilde{\mathbf{k}}}c^{}_{\tilde{\mathbf{m}},\tilde{\mathbf{k}} } \notag\\&\hspace{10pt}+(e^{-im_x b}-1) J^+_{yw}  c^\dagger_{\tilde{\mathbf{m}}-\mathbf{e}_y,\tilde{\mathbf{k}}  } c^{}_{\tilde{\mathbf{m}},\tilde{\mathbf{k}} }\,,
\label{eq:SSH}
\end{align}
and $J^\pm_{\mu\nu}~=~t_\mu~\pm~(-1)^{m_\mu}2t_{\mu\nu}\cos(\text{k}_\nu)$. 
Before applying dimensional reduction to the 4D ancestor model $\hat{H}^{4D}$, we first illustrate the (2D$\to $1D)-reduction of the 2D Creutz model $\hat{H}_{\mu\nu}(\tilde{\mathbf{m}},\tilde{\mathbf{k}}  )$ [cf.~Eq.~\eqref{eq:Creutz}] to the 1D Hamiltonian $\hat{h}_{\mu}(\text{k}_\nu)=\sum_{m_\mu}\hat{H}_{\mu\nu}(\tilde{\mathbf{m}},\tilde{\mathbf{k}}  )$, where $\text{k}_\nu$ is treated as an external parameter \footnote{for convenience we drop the dependence on the remaining coordinates $\text{m}_\sigma$ and $\text{k}_\rho$ since the Hamiltonian $\hat{h}_{\mu}(\text{k}_\nu)$ is independent of them}.
In accord with the magnetic unit cell imposed by the $\pi$-flux threading of the 2D Cruez model, the descendant Hamiltonian $\hat{h}_{\mu}(\text{k}_\nu)$ defines a 1D chain with a unit cell with two degrees of freedom, identical to the Su-Schrieffer-Heeger (SSH) chain~\cite{su1979solitons}. 
The topological invariant pertaining to the 1D SSH model is the bulk dipole moment $P_{\mu} ( \text{k}_\nu)$ (a.k.a. polarization), which can be calculated using the Wilson loop formalism~\cite{king1993theory,benalcazar2017quantized}. 
As a function of $\text{k}_\nu$, the bulk dipole $P_\mu (\text{k}_\nu) \,\text{mod}\,1$ is quantized to two values, $0$ and $1/2$, that are discontinuously connected at gap closing points.  
The quantisation of $P_{\mu} ( \text{k}_\nu)$ is imposed by the chiral symmetry (of both the 1D descendant and the 2D ancestor Creuz models) that ensures the existence of gap closing points in the $(\text{k}_\mu,\text{k}_\nu)$-parameter space, i.e., at the 2D Dirac cones [cf.~Fig.~1(b)]. 
An onsite staggered potential $ (-1)^{m_\mu } \mu_0  c^\dagger_{\mathbf{m}  }c^{}_{\mathbf{m} }$ (or similarly a nonzero hopping $t_\nu$ in the $\nu$-direction~\cite{hahaha} of the ancestor 2D Creutz model) breaks chiral symmetry and gaps the Dirac cone spectrum.
As a consequence, the two-dimensional $(\mathrm{{k}}_\mu,\mathrm{{k}}_\nu)$-parameter space acquires a well-defined \Fst Chern number $c_1 = \int_0 ^{2\pi} \text{dk}_\nu \partial_{\text{k}_\nu} P_\mu (\text{k}_\nu)$ given by the integral of the change of dipole moment over the entire descendant family of 1D models $\hat{h}_{\mu}(\text{k}_\nu)$, see Fig.~1(c)~\cite{thouless1983quantization,vanderbilt1993electric,lohse2016thouless}.
Hence, the adiabatic evolution of $\hat{h}_{\mu}(\text{k}_\nu (t))$ along a closed path realizes a dynamical version of the 2D QHE, dubbed topological pumping, where $c_1$ charges are transported across the 1D system per pump-cycle~\cite{thouless1983quantization, kraus2012topological,lohse2016thouless}.
Therefore, a nontrivial value $c_1 \ne 0$ means that the bulk dipole must wind as a function of $\text{k}_v$, cf.~Fig.~1(c).

%%%%%%%%%%%%%%%%%%%%%%%%%%%%%%%%%%%%%%%%%%%%%%%%%%%%%%%%%%%%%%%%%%%%%%%5
\begin{figure}[t!]
	\centering		
	\includegraphics[width=1\linewidth]{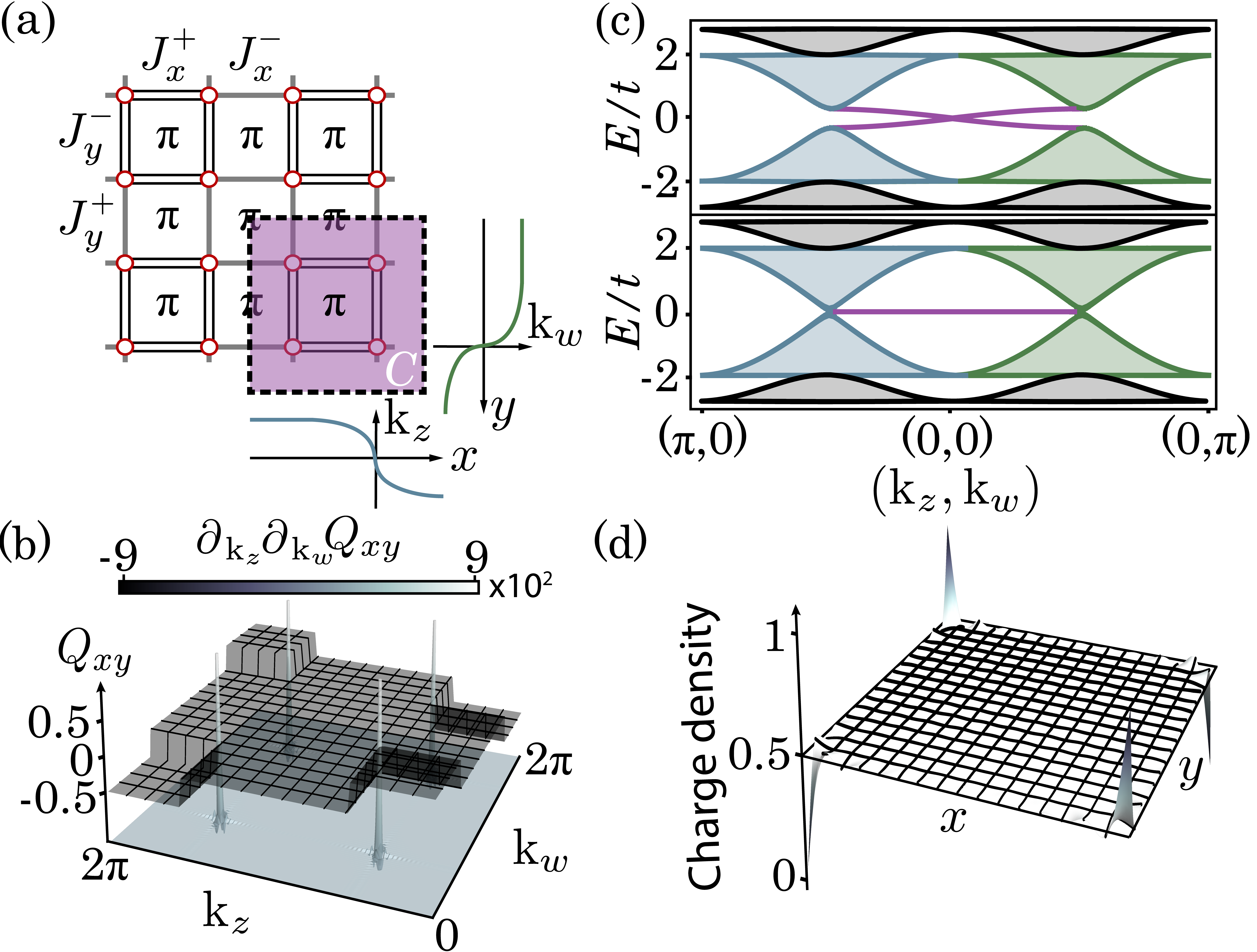}\label{fig2}	
	\caption{The 2D descendant family. 
		(a) The descendant model $\hat{h}^{\pi}_{xy}(\tilde{\mathbf{k}})$. %obtained by (4D$\to$2D)-dimensional reduction of Eq.~\eqref{HamProd} with $b=\pi$: A square lattice in the $xy$-plane made out of orthogonal SSH chains with alternating hopping in both $x$- and $y$-directions, threaded by a magnetic field with $\pi$ flux quanta per $xy$-plaquette. 
		Single (double) lines denote hopping amplitude $J^-_\mu$ ($J^+_\mu$). 
		The area $C$ enclosing the interface of the 2D material with the vacuum (dashed purple square), as well as the corresponding domain-wall configuration in the $\tilde{\mathbf{k}}$-parameter space are shown. 
		(b) The quadrupole moment $Q_{xy} (\tilde{\mathbf{k}})$ of $\hat{h}^{\pi}_{xy}(\tilde{\mathbf{k}})$ and the curvature $\partial_{\mathrm{k}_z}\partial_{\mathrm{k}_w}Q_{xy} (\tilde{\mathbf{k}})$ associated with the \Snd Chern number $c_2$; the total area under the latter is an integer. %of the ancestor 4D model $\sum_{\tilde{\mathbf{k}}} \hat{h}^{\pi}_{xy}(\tilde{\mathbf{k}})$.
		(c) The open boundary spectrum of $\hat{h}^{\pi}_{xy}(\tilde{\mathbf{k}})$, for a selected path in $\tilde{\mathbf{k}}$. Depicted are bulk bands (gray), upper/lower edge states (blue), right/left edge states (green), and corner states (purple). The top (bottom) spectrum has broken (preserved) chiral symmetry with $t_z=t_w=0.03$ ($t_z=t_w=0$). 
		The $\tilde{\mathbf{k}}$-parameter space of the latter is divided into a nontrivial region with zero-energy corner states and a trivial region with no zero-energy solutions.
		(d) The charge density of $\hat{h}^{\pi}_{xy}(\tilde{\mathbf{k}})$ at half filling and $\tilde{\mathbf{k}} = \left(0,0\right)$ has $\pm 1/2$ charge deviation at the corners. 
		In (b), (c) and (d), we used $t_x = t$, $t_x /t_y = 1$ and $t_{xz} /t_x=t_{yw} /t_y=0.45$. In (b) we used $t_z=t_w=0.001t$ to minimally open the gap.
	} 
\end{figure}
%%%%%%%%%%%%%%%%%%%%%%%%%%%%%%%%%%%%%%%%%%%%%%%%%%%%%%%%%%%%%%%%%%%%%%%5
%

A nonvanishing 1D bulk dipole $P_\mu (\text{k}_\nu)\neq 0$ has corresponding boundary effects, where 0D sub-gap states appear at the interface of the material with the vacuum, see Fig.~1(d)~\cite{vanderbilt1993electric}. 
In order to calculate the charge $q_S$ accumulated in a region $S$ enclosing the interface between two 1D insulators [see Fig.~1(e)], we linearise the dynamics around zero energy and obtain the low-energy Hamiltonian $\hat{h}(\text{k}_\nu)=\sum\limits_{|\text{k}_\mu|\le\Lambda}\mathbf{d}\cdot \bm{\sigma}$ (i.e., we use the massive Jakiw-Rebbi model~\cite{jackiw1976solitons}), where $\mathbf{d} = \{{v} (\mathrm{k}_\nu) \mathrm{k}_\mu, \mu_1 (\mathrm{k}_\nu), \mu_0(\mathrm{k}_\nu)\}$ is a real-valued vector, $\bm{\sigma} = \{\sigma_x , \sigma_y , \sigma_z\}$ are three matrices satisfying the Clifford algebra $\{\sigma_i ,\sigma_j\} = 2\delta_{ij} $, and $\Lambda$ is the cut-off energy scale of the low-energy theory. 
The resulting accumulated charge $q_S$ is
\begin{align}
q_S =\frac{1}{2\pi}\int_{\tilde{S}}\hat{\mathbf{\textbf{d}}}\cdot (\partial_{k_\mu}\hat{\textbf{\text{d}}} \times \partial_{r_{\mu}}\hat{\textbf{\text{d}}})\mathrm{dk}_{\mu}\mathrm{d}r_{\mu}\,,
\label{eq:charge of boundary 1D}
\end{align}
%where $\hat{\vec{d}}=\frac{\vec{d}}{|\vec{d}|}$, $\vec{d} = \{{v}_\mu (\mathrm{k}_\nu) \mathrm{k}_\mu, \delta_1 (\mathrm{k}_\nu), \mu_0\}$ is a real-valued vector, $v_\mu (\mathrm{k}_\nu)= J_{\mu\nu}^+ -J_{\mu\nu}^- $, $\delta_1 (\mathrm{k}_\nu)= J_{\mu\nu}^+ +J_{\mu\nu}^- $, 
where $\hat{\mathbf{d}}=\frac{\mathbf{d}}{|\mathbf{d}|}$, and $\tilde{S} = S\times [-\Lambda,\Lambda ]$ is the integration domain.
The interface between the two insulators in real space is equivalent to a domain wall in the $\mathrm{k}_\nu$-parameter space~\cite{vanderbilt1993electric}, see Fig.~1(e). Hence, the accumulated charge \eqref{eq:charge of boundary 1D} is the Berry flux (or \Fst Chern flux) attached to the corresponding region defined by $\tilde{S}$ in the 2D BZ of the ancestor Creuz Hamiltonian $\sum_{\text{k}_\nu} \hat{h}(\text{k}_\nu)$. 
In the limit where chiral symmetry is restored, $|q_S|$ becomes quantized to two values, $1/2$ or $0$, that correspond to encircling or not-encircling a singularity with $\pm 1/2$ Berry flux.     
Thus, the 1D family $\hat{h}_{\mu}(\text{k}_\nu)$ is divided into a trivial region with $q_S  = 0 $, and a nontrivial region with $|q_S | = 1/2 $, in accord with the value of the bulk dipole $P_\mu (\text{k}_\nu)$, cf.~Fig.~\ref{fig1}(c). 
This is known as the bulk-boundary correspondence of 1D TIs, i.e., the relation between the quantized topological index (bulk dipole) and charge at the boundary~\cite{graf2018bulk}. 
Note that Eq.~\eqref{eq:charge of boundary 1D} also describes the accumulation of nontopological charge at the boundary between two insulators (Tamm states), arising from surface polarizability~\cite{example}.

In similitude to the (2D$\to$1D)-reduction of a 2D chiral semimetal (Creuz model) to a 1D TI (SSH model), the main goal of this work is to demonstrate that the 4D Hamiltonian $\hat{H}^{4D}$ leads to 2D second-order TIs. 
A (4D$\to $2D)-dimensional reduction of Eq.~\eqref{HamProd} yields the 2D descendant family $\hat{h}^{b}_{xy}(\tilde{\mathbf{k}})=\sum_{\tilde{\mathbf{m}}}\left[\hat{H}_{xz}(\tilde{\mathbf{m}},\tilde{\mathbf{k}} )+\hat{H}_{yw}(\tilde{\mathbf{m}},\tilde{\mathbf{k}}) + \Delta\hat{H}^{b} _{xy} (\tilde{\mathbf{m}},\tilde{\mathbf{k}})\right]$, describing a square lattice in the $xy$-plane made out of SSH chains [see Eq.~\eqref{eq:Creutz}] in both the $x$- and $y$-directions, and where each $xy$-plaquette is threaded by a magnetic field with $b$ flux quanta, see Fig.~2(a).

%%%%%%%%%%%%%%%%%%%%%%%%%%%%%%%%%%%%%%%%%%%%%%%%%%%%%%%%%%%%%%%%%%%%%%%%%%%%%%%%%%%%%%%%%%%%%%%%%%%%%%%%%5
\begin{figure*}[t!]
	\centering		
	\includegraphics[width=1\linewidth]{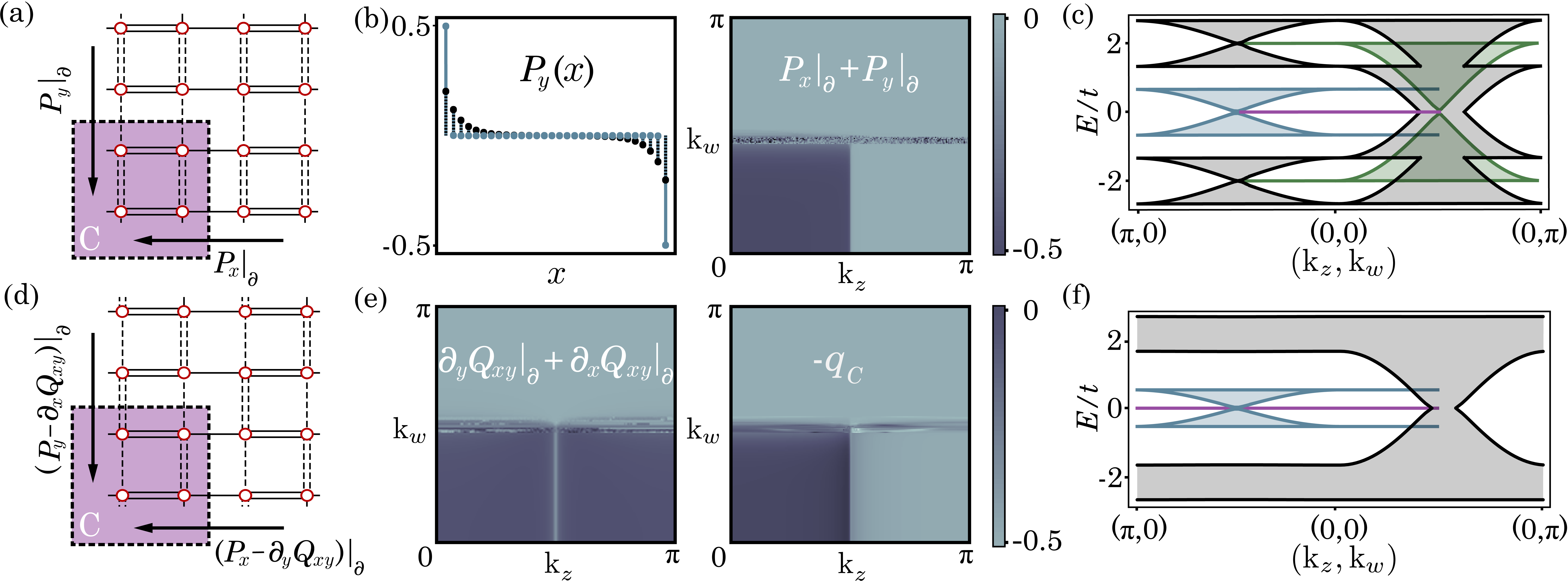}\label{fig3}	
	\caption{Descendant 2D models and their multipole description. 
		(a) The model $\hat{h}^{0}_{xy}(\tilde{\mathbf{k}})$. Single (double) lines denote a hopping strength of $J^-_\mu$ ($J^+_\mu$). Dashed (solid) lines denote $x$- ($y$-) hopping amplitudes. 
		Nonzero surface dipole moments $P_y |_{\partial}$ and $P_x |_{\partial}$ result in charge accumulation $q_C$ in a region $C$ enclosing the corner. 
		%		For the two plots we have used ${t_x / t_y} = {1/3}$ and $t_z = t_w = 0.003$. 
		(b) \textit{left:} The tangential dipole moment along $y$ as a function of lattice sites along $x$ of $\hat{h}^{0}_{xy}(\tilde{\mathbf{k}})$ for $\tilde{\mathbf{k}} = \left(0,0\right)$ (blue) and $\tilde{\mathbf{k}} = \left(5\pi/11,0\right)$ (black). 
		As the edge-gap closing point $\tilde{\mathbf{k}} = \left(\pi/2,0\right)$ is approached, the localisation length of the surface dipoles diverges. 
		\textit{right:} The charge $q_C$ as a function of $\tilde{\mathbf{k}}$, showing two distinct regions with $q_C = 0$ and $q_C = 1/2$. 
		%The latter arises due to the combination of converging surface dipole moments $P_y |_{\partial}$ and $P_x |_{\partial}$. 
		Discontinuities around $\mathrm{k}_w = \pi/2$ are due to bulk bands approaching zero energy in [cf.~(c)]. 
		(c) The open boundary spectrum of $\hat{h}^{0}_{xy}(\tilde{\mathbf{k}})$ for a selected path in $\tilde{\mathbf{k}}$. 
		Gap closings occur at the bulk- or edge-spectra, dividing the $\tilde{\mathbf{k}}$-parameter space into regions with/out zero-energy corner states.
		%The $\tilde{\mathbf{k}}$-parameter space is divided into regions indexed by the appearance of zero-energy corner states, while phase transitions happen at bulk- or edge-closing points.
		(d) The model $\hat{h}^{\pi,\pi}_{xy}(\tilde{\mathbf{k}})$.
		Lines denote hopping amplitudes as in (a).
		For this model, the charge $q_C$ arises due to nonzero surface dipole moments $P_y |_{\partial}$ and $P_x |_{\partial}$, as well as nonzero quadrupole moments $\partial_{x}Q_{xy} |_{\partial}$ and $\partial_{y}Q_{xy}  |_{\partial}$. 
		(e) \textit{left:} The contribution to $q_C$ from the quadrupole moments $\partial_{x}Q_{xy} |_{\partial}$ and $\partial_{y}Q_{xy}  |_{\partial}$.
		\textit{right:} The charge $q_C$ as a function of $\tilde{\mathbf{k}}$.
		%In the latter case, the charge accumulation arise due to the combination of nonzero surface dipole moments $P_y |_{\partial}$ and $P_x |_{\partial}$ and quadrupole moments $\partial_{x}Q_{xy} |_{\partial}$ and $\partial_{y}Q_{xy}  |_{\partial}$.
		(f)  The open boundary spectrum of $\hat{h}^{\pi,\pi}_{xy}(\tilde{\mathbf{k}})$ for a selected path in $\tilde{\mathbf{k}}$, exhibiting regions with: (i) zero-energy states localised at the upper/lower left corner, (ii) zero-energy states localised at the upper/lower right corner, and (iii) no zero-energy solutions. 
		In (b), (c), (e) and (f), we have used $t_x=t$, $t_y /t_x = 1/10$, $t_{xz} /t_x=t_{yw} /t_y=0.45$, and an onsite staggered mass $\mu_0=0.001$.} 
\end{figure*}
%%%%%%%%%%%%%%%%%%%%%%%%%%%%%%%%%%%%%%%%%%%%%%%%%%%%%%%%%%%%%%%%%%%%%%%%%%%%%%%%%%%%%%%%%%%%%%%%%%%%%%%5
We first consider the case $b=\pi$.
The topological invariant of the resulting 2D descendant model is associated with a bulk quadrupole moment $Q_{xy}$ that can be calculated using nested Wilson loops~\cite{benalcazar2017quantized,benalcazar2017electric,serra2018observation}, see Fig.~2(b). 
As a function of $\tilde{\mathbf{k}}$, the bulk quadrupole $Q_{xy}$ takes quantized values, either $0$ or $\pm\frac{1}{2}$~\cite{benalcazar2017quantized}. 
Breaking chiral symmetry with an onsite potential $ (-1)^{m_x + m_y} \mu_0 $ (or similarly when nearest-neighbor hopping amplitudes $t_z\ne0\ne t_w  $ appear in the $z$- and $w$- direction of the ancestor 4D model) makes the \Snd Chern number of the $(\mathrm{k}_x , \mathrm{k}_y , \mathrm{k}_z ,\mathrm{k}_w)$-parameter space well-defined and equal~\cite{appendix} to $	c_2 = \int_{\mathds{T}^2}  \partial_{\mathrm{k}_z} \partial_{\mathrm{k}_w} Q_{xy}(\tilde{\mathbf{k}})\text{d}^2 \tilde{\mathbf{k}} $, see Fig.~2(b). 
Hence, the adiabatic evolution of $\hat{h}^{\pi}_{xy}(\tilde{\mathbf{k}}(t))$ over a closed surface in the $\tilde{\mathbf{k}}$-parameter space realizes a dynamical version of the 4D QHE, called 2D topological pumping~\cite{kraus2013four,lohse2018exploring,zilberberg2018photonic}, where charge proportional to $c_2$ is transported across the 2D system per pump cycle. 
Therefore, a nontrivial value of $c_2\ne0$ results in the ``winding" of the bulk quadrupole as a function $\tilde{\mathbf{k}}$, see Fig.~2(b). 
The bulk responses of the descendant 2D pump $\hat{h}^{\pi}_{xy}(\tilde{\mathbf{k}})$ have associated boundary phenomena~\cite{zilberberg2018photonic} where: (i) 1D edge states, i.e., states localized in one of the two dimensions but extended in the other, appear in the spectrum, and (ii) sub-gap 0D corner states, i.e., states localized in both dimensions, disperse as a function of $\tilde{\mathbf{k}}$, see Figs.~2(c) and (d).

For a generic 2D material with low-energy Hamiltonian $\hat{h}(\tilde{\mathbf{k}})=\int_{|{\mathbf{k}}|\le\Lambda}\mathbf{d}\cdot \mathbf{\Gamma}\text{d}^2\text{k}$, where $\mathbf{d}=\{\mu_0 (\tilde{\mathbf{k}}), \linebreak\mu_1 (\tilde{\mathbf{k}}), \mu_2 (\tilde{\mathbf{k}}), {v}_x (\tilde{\mathbf{k}}) \mathrm{k}_x, {v}_y (\tilde{\mathbf{k}}) \mathrm{k}_y  \}$ is a real-valued vector, $\mathbf{\Gamma}=\{\Gamma_0,...,\Gamma_4\}$ are five anticommuting matrices $\{\Gamma_\mu , \Gamma_\nu\} = 2\delta_{\mu\nu}$, and $\Lambda$ is a cut-off energy scale, we derive~\cite{appendix} the charge accumulation $q_C$ in a region $C$ enclosing the corner of the system [cf.~Fig.~2(a)]
\begin{align}
	q_C =\int_{\tilde{C}}\hat{\mathbf{\textbf{d}}}\cdot (\partial_{\mathrm{k}_x}\hat{\textbf{\text{d}}} \times \partial_{\mathrm{k}_y}\hat{\mathbf{\textbf{d}}}\times \partial_{x}\hat{\mathbf{\textbf{d}}} \times \partial_{y}\hat{\mathbf{\textbf{d}}})\mathrm{d}^2 \mathrm{{k}}\mathrm{d}^2 \mathrm{r}\,,
	\label{eq:charge of boundary 2D}
\end{align}
where $\hat{\mathbf{d}}=\frac{\mathbf{d}}{|\mathbf{d}|}$, and $\tilde{C} = C\times [-\Lambda,\Lambda ]^2$ is the integration domain.
Since the corner of the material can be expressed as the intersection of two domain walls in the $\tilde{\mathbf{k}}$-parameter space [cf.~Fig.~2(a)], $q_C$ is equivalent to the \Snd Chern flux attached to the region defined by $\tilde{C}$ in the 4D BZ of the ancestor Hamiltonian \eqref{HamProd}. %$\sum_{\tilde{\mathbf{k}}}\hat{h}(\tilde{\mathbf{k}})$. 
In the limit where chiral symmetry is restored, $|q_C|$ becomes quantised to $0$ or $1/2$, corresponding to encircling or not-encircling a 4D singularity with $\pm \frac{1}{2}$ \Snd Chern flux. 
For the 2D family $\hat{h}^{\pi}_{xy}(\tilde{\mathbf{k}})$, we find that the $\tilde{\mathbf{k}}$-parameter space is divided into trivial regions with $|q_C| = 0$ and a nontrivial region with $|q_C| = 1/2$; the latter exhibiting zero-energy states localized at the corners, in accord with the value of the bulk quadrupole $Q_{xy}$, cf.~Figs.~2(b)-(d). 

The connection between charge accumulation at the 2D corner and the \Snd Chern flux is a key outcome of this work. 
However, such charges can arise due to bulk topology as well as due to boundary effects. 
In general, the charge accumulation $q_C$ in a region $C$ enclosing the corner of a finite-sized macroscopic 2D material interfaced with another material can be calculated using the electric multipole expansion~\cite{benalcazar2017quantized}
\begin{align}
q_C = \int_{C} (\rho_{\text{bulk}} + \rho_{\partial} + \rho_{\partial \partial} )\mathrm{d}^2\mathbf{r}\,,
\end{align}
where $\rho_{\text{bulk}}=  - \nabla\cdot\vec{P} + \frac{1}{2}\partial_i \partial_j Q_{ij}$ are the contributions due to the bulk dipole $\vec{P}$ and quadrupole $Q_{xy}$ densities, 
$\rho_{\partial}= \hat{\vec{n}}\cdot \vec{P}|_{\partial }  -\hat{n}_i \partial_j  Q_{ij}|_{\partial }$ are the contributions due to a ``free" edge dipole $ \hat{\vec{n}}\cdot \vec{P}|_{\partial }$ and quadrupole $\hat{n}_i \partial_j  Q_{ij}|_{\partial }$ densities, 
and $\rho_{\partial \partial} =\frac{1}{2}\hat{n}^{\alpha}_i \hat{n}^{\beta}_j Q_{ij}$ is the contribution due to a point charge created by a ``free" quadrupole density at the intersection of two edges with normal vectors $\hat{\vec{n}}^{\alpha}$ and $\hat{\vec{n}}^{\beta}$. 
Hence, a nontrivial value of $q_C$ can also originate from a combination of bulk and surface terms~\cite{appendix}. 

For the 2D Hamiltonian $\hat{h}^{\pi}_{xy}(\tilde{\mathbf{k}})$, the only nonvanishing contributions to the corner charge $q_C $ arise from a quantised bulk quadrupole $Q_{xy}$.
On the other hand, starting from the 4D ancestor model Eq.~\eqref{HamProd} with $b=0$ we obtain, upon (4D$\to $2D)-dimensional reduction, a 2D tight-binding model $\hat{h}^{0}_{xy}(\tilde{\mathbf{k}})$ of SSH chains $\hat{h}_{x}(\text{k}_z)$ along the $x$-direction coupled to SSH chains $\hat{h}_{y}(\text{k}_w )$ in the $y$-direction, see Fig.~3(a). 
The descendant 2D family has zero bulk quadrupole density $Q_{xy}$ but nonzero edge dipole densities $\vec{P}|_\partial$ that result in two distinct phases with $q_C =0$ or $1/2$, see Fig.~3(b). 
In the latter, zero-energy states appear at the corners, while phase transitions happen at (bulk- or edge-) gap closing points, see Fig.~3(c). 
As a third example, we start from the 4D ancestor model $\hat{H}^{4D}$ with $b=0$ and thread a $\pi$-flux through the $xw$-plane.
This leads, upon (4D$\to $2D)-dimensional reduction, to a 2D descendant family denoted by $\hat{h}^{\pi,\pi}_{xy}(\tilde{\mathbf{k}})$ and described by SSH chains $\hat{h}_{x}(\text{k}_z)$ along the $x$-direction, coupled to alternating SSH chains $\hat{h}_{y}(\text{k}_w +\pi x)$ in the $y$-direction, see Fig.~3(d). 
The charge accumulation $q_C$ is now a combination of bulk and surface terms that sum to quantised values, $0$ or $1/2$, see Fig.~3(e).
The spectrum is, once more, separated into regions characterized by the appearance of zero-energy states, while phase transitions happen at (bulk- or edge-) gap closing points, see Fig.~3(f). 
In all three cases, the charge accumulation at the intersection of two edges is associated to a nontrivial value of the \Snd Chern flux attached to the region defined by $C$ in the 4D BZ of the ancestor Hamiltonian~\cite{appendix}.

The relationship between the 4D chiral semimetal $\hat{H}^{4D}$ and 2D second-order TIs offers a plethora of generalizations. 
Namely, there is a wide variety of 4D ancestor models that can be constructed where various planes are threaded with $2\pi/q$ fluxes (with $q$ an even integer) and different directions are dimensionally reduced. 
Moreover, our methodology is readily extended to a (6D$\to $3D)-reduction where the charge accumulated at the corner (i.e., at the intersection of three edges) is associated to a 3$^{\text{rd}}$ Chern number~\cite{petrides2018six}. 
Such charges can arise from a combination of octapole, quadrupole and dipole moments.
Equivalently, our procedure offers multiple topological pump realisations, where a time-dependent adiabatic evolution results in charge transport across the descendant system, in response to the modulation of the bulk dipole, quadrupole and octapole moments; this naturally explains the appearance of surface, hinge and corner modes [cf. Fig.~2(c) and Ref. [24]]. 
   
% We have already shown the most-studied 2D high-order TIs are obtained using this procedure~\cite{} [the SSHxSSH model with its dual mirror symmetries~\cite{} and the $C_4$ symmetry model with the $\pi$ flux in the plane~\cite{}], as well as a new one model that has nonsymmorphic symmetries with only part of its corners having quantized charges, see Fig.~3(?). Our methodology is readily extended with dimensional reduction from other Chern insulators, e.g., $6D\rightarrow 3D$, where a relation between an octapole and a 3$^{\text{rd}}$ Chern number can be derived~\cite{petrides2018}. Similarly, the dispersion associated with the bulk responses in Chern insulators directly explains the dispaersion of hinge modes~\cite{zilberbergphotons}.

In this paper, we find that dimensional reduction reveals a connection between the physics of high-order TIs, topological pumps and Chern insulators. As an example, we show that the (4D$\to$2D)-dimensional reduction of a 4D Chern insulator results into different families of 2D topological pumps. By taking the limit where the former becomes chiral, we obtain a relation between the corner charge found in 2D topological pumps and 2D second-order TIs.
We use a low-energy continuum theory to calculate the charge accumulation at the intersection of two edges and associate it to a nontrivial value of the \Snd Chern flux. 
In the high-energy description the corner charge arises from a combination of bulk and surface multipole moments.
Hence, the definition of an invariant associated to the charge accumulation at the corner of a high-order TIs can be readily obtained from high-dimensional Chern forms and various models can be derived from a single ancestor high-dimensional insulator, using dimensional reduction. 

We thank H.~M.~Price, I.~Mondragon, M.~Soriente, T.~Wolf, and M.~Rechtsman for fruitful discussions. We acknowledge financial support from the Swiss national science foundation.

\bibliography{QuadRef.bib}

\begin{thebibliography}{49}%
\makeatletter
\providecommand \@ifxundefined [1]{%
 \@ifx{#1\undefined}
}%
\providecommand \@ifnum [1]{%
 \ifnum #1\expandafter \@firstoftwo
 \else \expandafter \@secondoftwo
 \fi
}%
\providecommand \@ifx [1]{%
 \ifx #1\expandafter \@firstoftwo
 \else \expandafter \@secondoftwo
 \fi
}%
\providecommand \natexlab [1]{#1}%
\providecommand \enquote  [1]{``#1''}%
\providecommand \bibnamefont  [1]{#1}%
\providecommand \bibfnamefont [1]{#1}%
\providecommand \citenamefont [1]{#1}%
\providecommand \href@noop [0]{\@secondoftwo}%
\providecommand \href [0]{\begingroup \@sanitize@url \@href}%
\providecommand \@href[1]{\@@startlink{#1}\@@href}%
\providecommand \@@href[1]{\endgroup#1\@@endlink}%
\providecommand \@sanitize@url [0]{\catcode `\\12\catcode `\$12\catcode
  `\&12\catcode `\#12\catcode `\^12\catcode `\_12\catcode `\%12\relax}%
\providecommand \@@startlink[1]{}%
\providecommand \@@endlink[0]{}%
\providecommand \url  [0]{\begingroup\@sanitize@url \@url }%
\providecommand \@url [1]{\endgroup\@href {#1}{\urlprefix }}%
\providecommand \urlprefix  [0]{URL }%
\providecommand \Eprint [0]{\href }%
\providecommand \doibase [0]{http://dx.doi.org/}%
\providecommand \selectlanguage [0]{\@gobble}%
\providecommand \bibinfo  [0]{\@secondoftwo}%
\providecommand \bibfield  [0]{\@secondoftwo}%
\providecommand \translation [1]{[#1]}%
\providecommand \BibitemOpen [0]{}%
\providecommand \bibitemStop [0]{}%
\providecommand \bibitemNoStop [0]{.\EOS\space}%
\providecommand \EOS [0]{\spacefactor3000\relax}%
\providecommand \BibitemShut  [1]{\csname bibitem#1\endcsname}%
\let\auto@bib@innerbib\@empty
%</preamble>
\bibitem [{\citenamefont {Hasan}\ and\ \citenamefont {Kane}(2010)}]{RMP_TI}%
  \BibitemOpen
  \bibfield  {author} {\bibinfo {author} {\bibfnamefont {M.~Z.}\ \bibnamefont
  {Hasan}}\ and\ \bibinfo {author} {\bibfnamefont {C.~L.}\ \bibnamefont
  {Kane}},\ }\href@noop {} {\bibfield  {journal} {\bibinfo  {journal} {Rev.
  Mod. Phys.}\ }\textbf {\bibinfo {volume} {82}},\ \bibinfo {pages} {3045}
  (\bibinfo {year} {2010})}\BibitemShut {NoStop}%
\bibitem [{\citenamefont {Qi}\ and\ \citenamefont
  {Zhang}(2011{\natexlab{a}})}]{RMP_TI2}%
  \BibitemOpen
  \bibfield  {author} {\bibinfo {author} {\bibfnamefont {X.-L.}\ \bibnamefont
  {Qi}}\ and\ \bibinfo {author} {\bibfnamefont {S.-C.}\ \bibnamefont {Zhang}},\
  }\href@noop {} {\bibfield  {journal} {\bibinfo  {journal} {Rev. Mod. Phys.}\
  }\textbf {\bibinfo {volume} {83}},\ \bibinfo {pages} {1057} (\bibinfo {year}
  {2011}{\natexlab{a}})}\BibitemShut {NoStop}%
\bibitem [{\citenamefont {Ozawa}\ \emph {et~al.}(2018)\citenamefont {Ozawa},
  \citenamefont {Price}, \citenamefont {Amo}, \citenamefont {Goldman},
  \citenamefont {Hafezi}, \citenamefont {Lu}, \citenamefont {Rechtsman},
  \citenamefont {Schuster}, \citenamefont {Simon}, \citenamefont {Zilberberg}
  \emph {et~al.}}]{ozawareview}%
  \BibitemOpen
  \bibfield  {author} {\bibinfo {author} {\bibfnamefont {T.}~\bibnamefont
  {Ozawa}}, \bibinfo {author} {\bibfnamefont {H.~M.}\ \bibnamefont {Price}},
  \bibinfo {author} {\bibfnamefont {A.}~\bibnamefont {Amo}}, \bibinfo {author}
  {\bibfnamefont {N.}~\bibnamefont {Goldman}}, \bibinfo {author} {\bibfnamefont
  {M.}~\bibnamefont {Hafezi}}, \bibinfo {author} {\bibfnamefont
  {L.}~\bibnamefont {Lu}}, \bibinfo {author} {\bibfnamefont {M.}~\bibnamefont
  {Rechtsman}}, \bibinfo {author} {\bibfnamefont {D.}~\bibnamefont {Schuster}},
  \bibinfo {author} {\bibfnamefont {J.}~\bibnamefont {Simon}}, \bibinfo
  {author} {\bibfnamefont {O.}~\bibnamefont {Zilberberg}},  \emph {et~al.},\
  }\href@noop {} {\bibfield  {journal} {\bibinfo  {journal} {arXiv preprint
  arXiv:1802.04173}\ } (\bibinfo {year} {2018})}\BibitemShut {NoStop}%
\bibitem [{\citenamefont {Kitaev}(2009)}]{kitaev2009periodic}%
  \BibitemOpen
  \bibfield  {author} {\bibinfo {author} {\bibfnamefont {A.}~\bibnamefont
  {Kitaev}},\ }in\ \href@noop {} {\emph {\bibinfo {booktitle} {AIP conference
  proceedings}}},\ Vol.\ \bibinfo {volume} {1134}\ (\bibinfo {organization}
  {AIP},\ \bibinfo {year} {2009})\ pp.\ \bibinfo {pages} {22--30}\BibitemShut
  {NoStop}%
\bibitem [{\citenamefont {Ryu}\ \emph {et~al.}(2010)\citenamefont {Ryu},
  \citenamefont {Schnyder}, \citenamefont {Furusaki},\ and\ \citenamefont
  {Ludwig}}]{ryu2010topological}%
  \BibitemOpen
  \bibfield  {author} {\bibinfo {author} {\bibfnamefont {S.}~\bibnamefont
  {Ryu}}, \bibinfo {author} {\bibfnamefont {A.~P.}\ \bibnamefont {Schnyder}},
  \bibinfo {author} {\bibfnamefont {A.}~\bibnamefont {Furusaki}}, \ and\
  \bibinfo {author} {\bibfnamefont {A.~W.}\ \bibnamefont {Ludwig}},\
  }\href@noop {} {\bibfield  {journal} {\bibinfo  {journal} {New Journal of
  Physics}\ }\textbf {\bibinfo {volume} {12}},\ \bibinfo {pages} {065010}
  (\bibinfo {year} {2010})}\BibitemShut {NoStop}%
\bibitem [{\citenamefont {Altland}\ and\ \citenamefont
  {Zirnbauer}(1997)}]{altland1997nonstandard}%
  \BibitemOpen
  \bibfield  {author} {\bibinfo {author} {\bibfnamefont {A.}~\bibnamefont
  {Altland}}\ and\ \bibinfo {author} {\bibfnamefont {M.~R.}\ \bibnamefont
  {Zirnbauer}},\ }\href@noop {} {\bibfield  {journal} {\bibinfo  {journal}
  {Physical Review B}\ }\textbf {\bibinfo {volume} {55}},\ \bibinfo {pages}
  {1142} (\bibinfo {year} {1997})}\BibitemShut {NoStop}%
\bibitem [{\citenamefont {Fu}(2011)}]{fu2011topological}%
  \BibitemOpen
  \bibfield  {author} {\bibinfo {author} {\bibfnamefont {L.}~\bibnamefont
  {Fu}},\ }\href@noop {} {\bibfield  {journal} {\bibinfo  {journal} {Physical
  Review Letters}\ }\textbf {\bibinfo {volume} {106}},\ \bibinfo {pages}
  {106802} (\bibinfo {year} {2011})}\BibitemShut {NoStop}%
\bibitem [{\citenamefont {Kremer}\ \emph {et~al.}(2018)\citenamefont {Kremer},
  \citenamefont {Petrides}, \citenamefont {Meyer}, \citenamefont {Heinrich},
  \citenamefont {Zilberberg},\ and\ \citenamefont {Szameit}}]{kremer2018non}%
  \BibitemOpen
  \bibfield  {author} {\bibinfo {author} {\bibfnamefont {M.}~\bibnamefont
  {Kremer}}, \bibinfo {author} {\bibfnamefont {I.}~\bibnamefont {Petrides}},
  \bibinfo {author} {\bibfnamefont {E.}~\bibnamefont {Meyer}}, \bibinfo
  {author} {\bibfnamefont {M.}~\bibnamefont {Heinrich}}, \bibinfo {author}
  {\bibfnamefont {O.}~\bibnamefont {Zilberberg}}, \ and\ \bibinfo {author}
  {\bibfnamefont {A.}~\bibnamefont {Szameit}},\ }\href@noop {} {\bibfield
  {journal} {\bibinfo  {journal} {arXiv preprint arXiv:1805.05209}\ } (\bibinfo
  {year} {2018})}\BibitemShut {NoStop}%
\bibitem [{\citenamefont {Kraus}\ and\ \citenamefont
  {Zilberberg}(2012)}]{kraus2012topological}%
  \BibitemOpen
  \bibfield  {author} {\bibinfo {author} {\bibfnamefont {Y.~E.}\ \bibnamefont
  {Kraus}}\ and\ \bibinfo {author} {\bibfnamefont {O.}~\bibnamefont
  {Zilberberg}},\ }\href@noop {} {\bibfield  {journal} {\bibinfo  {journal}
  {Physical review letters}\ }\textbf {\bibinfo {volume} {109}},\ \bibinfo
  {pages} {116404} (\bibinfo {year} {2012})}\BibitemShut {NoStop}%
\bibitem [{\citenamefont {Kraus}\ and\ \citenamefont
  {Zilberberg}(2016)}]{kraus2016quasiperiodicity}%
  \BibitemOpen
  \bibfield  {author} {\bibinfo {author} {\bibfnamefont {Y.~E.}\ \bibnamefont
  {Kraus}}\ and\ \bibinfo {author} {\bibfnamefont {O.}~\bibnamefont
  {Zilberberg}},\ }\href@noop {} {\bibfield  {journal} {\bibinfo  {journal}
  {Nature Physics}\ }\textbf {\bibinfo {volume} {12}},\ \bibinfo {pages} {624}
  (\bibinfo {year} {2016})}\BibitemShut {NoStop}%
\bibitem [{\citenamefont {Bellissard}\ \emph {et~al.}(2000)\citenamefont
  {Bellissard}, \citenamefont {Herrmann},\ and\ \citenamefont
  {Zarrouati}}]{bellissard2000hull}%
  \BibitemOpen
  \bibfield  {author} {\bibinfo {author} {\bibfnamefont {J.}~\bibnamefont
  {Bellissard}}, \bibinfo {author} {\bibfnamefont {D.}~\bibnamefont
  {Herrmann}}, \ and\ \bibinfo {author} {\bibfnamefont {M.}~\bibnamefont
  {Zarrouati}},\ }\href@noop {} {\bibfield  {journal} {\bibinfo  {journal}
  {Directions in mathematical quasicrystals}\ }\textbf {\bibinfo {volume}
  {13}},\ \bibinfo {pages} {207} (\bibinfo {year} {2000})}\BibitemShut
  {NoStop}%
\bibitem [{\citenamefont {Shiozaki}(2017)}]{shiozaki2017k}%
  \BibitemOpen
  \bibfield  {author} {\bibinfo {author} {\bibfnamefont {K.}~\bibnamefont
  {Shiozaki}},\ }\href@noop {} {\bibfield  {journal} {\bibinfo  {journal}
  {Phys. Rev. B}\ }\textbf {\bibinfo {volume} {95}},\ \bibinfo {pages} {235425}
  (\bibinfo {year} {2017})}\BibitemShut {NoStop}%
\bibitem [{\citenamefont {Alexandradinata}\ \emph {et~al.}(2016)\citenamefont
  {Alexandradinata}, \citenamefont {Wang},\ and\ \citenamefont
  {Bernevig}}]{alexandradinata2016topological}%
  \BibitemOpen
  \bibfield  {author} {\bibinfo {author} {\bibfnamefont {A.}~\bibnamefont
  {Alexandradinata}}, \bibinfo {author} {\bibfnamefont {Z.}~\bibnamefont
  {Wang}}, \ and\ \bibinfo {author} {\bibfnamefont {B.~A.}\ \bibnamefont
  {Bernevig}},\ }\href@noop {} {\bibfield  {journal} {\bibinfo  {journal}
  {Physical Review X}\ }\textbf {\bibinfo {volume} {6}},\ \bibinfo {pages}
  {021008} (\bibinfo {year} {2016})}\BibitemShut {NoStop}%
\bibitem [{\citenamefont {Bellissard}(1992)}]{bellissard1992gap}%
  \BibitemOpen
  \bibfield  {author} {\bibinfo {author} {\bibfnamefont {J.}~\bibnamefont
  {Bellissard}},\ }in\ \href@noop {} {\emph {\bibinfo {booktitle} {From number
  theory to physics}}}\ (\bibinfo  {publisher} {Springer},\ \bibinfo {year}
  {1992})\ pp.\ \bibinfo {pages} {538--630}\BibitemShut {NoStop}%
\bibitem [{\citenamefont {Prodan}(2015)}]{prodan2015virtual}%
  \BibitemOpen
  \bibfield  {author} {\bibinfo {author} {\bibfnamefont {E.}~\bibnamefont
  {Prodan}},\ }\href@noop {} {\bibfield  {journal} {\bibinfo  {journal}
  {Physical Review B}\ }\textbf {\bibinfo {volume} {91}},\ \bibinfo {pages}
  {245104} (\bibinfo {year} {2015})}\BibitemShut {NoStop}%
\bibitem [{\citenamefont {Chiu}\ \emph {et~al.}(2016)\citenamefont {Chiu},
  \citenamefont {Teo}, \citenamefont {Schnyder},\ and\ \citenamefont
  {Ryu}}]{chiu2016classification}%
  \BibitemOpen
  \bibfield  {author} {\bibinfo {author} {\bibfnamefont {C.-K.}\ \bibnamefont
  {Chiu}}, \bibinfo {author} {\bibfnamefont {J.~C.}\ \bibnamefont {Teo}},
  \bibinfo {author} {\bibfnamefont {A.~P.}\ \bibnamefont {Schnyder}}, \ and\
  \bibinfo {author} {\bibfnamefont {S.}~\bibnamefont {Ryu}},\ }\href@noop {}
  {\bibfield  {journal} {\bibinfo  {journal} {Reviews of Modern Physics}\
  }\textbf {\bibinfo {volume} {88}},\ \bibinfo {pages} {035005} (\bibinfo
  {year} {2016})}\BibitemShut {NoStop}%
\bibitem [{\citenamefont {Qi}\ and\ \citenamefont
  {Zhang}(2011{\natexlab{b}})}]{qi2011topological}%
  \BibitemOpen
  \bibfield  {author} {\bibinfo {author} {\bibfnamefont {X.-L.}\ \bibnamefont
  {Qi}}\ and\ \bibinfo {author} {\bibfnamefont {S.-C.}\ \bibnamefont {Zhang}},\
  }\href@noop {} {\bibfield  {journal} {\bibinfo  {journal} {Reviews of Modern
  Physics}\ }\textbf {\bibinfo {volume} {83}},\ \bibinfo {pages} {1057}
  (\bibinfo {year} {2011}{\natexlab{b}})}\BibitemShut {NoStop}%
\bibitem [{\citenamefont {Thouless}\ \emph {et~al.}(1982)\citenamefont
  {Thouless}, \citenamefont {Kohmoto}, \citenamefont {Nightingale},\ and\
  \citenamefont {den Nijs}}]{thouless1982quantized}%
  \BibitemOpen
  \bibfield  {author} {\bibinfo {author} {\bibfnamefont {D.~J.}\ \bibnamefont
  {Thouless}}, \bibinfo {author} {\bibfnamefont {M.}~\bibnamefont {Kohmoto}},
  \bibinfo {author} {\bibfnamefont {M.~P.}\ \bibnamefont {Nightingale}}, \ and\
  \bibinfo {author} {\bibfnamefont {M.}~\bibnamefont {den Nijs}},\ }\href@noop
  {} {\bibfield  {journal} {\bibinfo  {journal} {Physical review letters}\
  }\textbf {\bibinfo {volume} {49}},\ \bibinfo {pages} {405} (\bibinfo {year}
  {1982})}\BibitemShut {NoStop}%
\bibitem [{\citenamefont {Thouless}(1983)}]{thouless1983quantization}%
  \BibitemOpen
  \bibfield  {author} {\bibinfo {author} {\bibfnamefont {D.}~\bibnamefont
  {Thouless}},\ }\href@noop {} {\bibfield  {journal} {\bibinfo  {journal}
  {Physical Review B}\ }\textbf {\bibinfo {volume} {27}},\ \bibinfo {pages}
  {6083} (\bibinfo {year} {1983})}\BibitemShut {NoStop}%
\bibitem [{\citenamefont {Kraus}(2012)}]{kraus2012ye}%
  \BibitemOpen
  \bibfield  {author} {\bibinfo {author} {\bibfnamefont {Y.}~\bibnamefont
  {Kraus}},\ }\href@noop {} {\bibfield  {journal} {\bibinfo  {journal} {Phys.
  Rev. Lett.}\ }\textbf {\bibinfo {volume} {109}},\ \bibinfo {pages} {106402}
  (\bibinfo {year} {2012})}\BibitemShut {NoStop}%
\bibitem [{\citenamefont {Verbin}\ \emph {et~al.}(2015)\citenamefont {Verbin},
  \citenamefont {Zilberberg}, \citenamefont {Lahini}, \citenamefont {Kraus},\
  and\ \citenamefont {Silberberg}}]{verbin2015topological}%
  \BibitemOpen
  \bibfield  {author} {\bibinfo {author} {\bibfnamefont {M.}~\bibnamefont
  {Verbin}}, \bibinfo {author} {\bibfnamefont {O.}~\bibnamefont {Zilberberg}},
  \bibinfo {author} {\bibfnamefont {Y.}~\bibnamefont {Lahini}}, \bibinfo
  {author} {\bibfnamefont {Y.~E.}\ \bibnamefont {Kraus}}, \ and\ \bibinfo
  {author} {\bibfnamefont {Y.}~\bibnamefont {Silberberg}},\ }\href@noop {}
  {\bibfield  {journal} {\bibinfo  {journal} {Physical Review B}\ }\textbf
  {\bibinfo {volume} {91}},\ \bibinfo {pages} {064201} (\bibinfo {year}
  {2015})}\BibitemShut {NoStop}%
\bibitem [{\citenamefont {Lohse}\ \emph {et~al.}(2016)\citenamefont {Lohse},
  \citenamefont {Schweizer}, \citenamefont {Zilberberg}, \citenamefont
  {Aidelsburger},\ and\ \citenamefont {Bloch}}]{lohse2016thouless}%
  \BibitemOpen
  \bibfield  {author} {\bibinfo {author} {\bibfnamefont {M.}~\bibnamefont
  {Lohse}}, \bibinfo {author} {\bibfnamefont {C.}~\bibnamefont {Schweizer}},
  \bibinfo {author} {\bibfnamefont {O.}~\bibnamefont {Zilberberg}}, \bibinfo
  {author} {\bibfnamefont {M.}~\bibnamefont {Aidelsburger}}, \ and\ \bibinfo
  {author} {\bibfnamefont {I.}~\bibnamefont {Bloch}},\ }\href@noop {}
  {\bibfield  {journal} {\bibinfo  {journal} {Nature Physics}\ }\textbf
  {\bibinfo {volume} {12}},\ \bibinfo {pages} {350} (\bibinfo {year}
  {2016})}\BibitemShut {NoStop}%
\bibitem [{\citenamefont {Kraus}\ \emph {et~al.}(2013)\citenamefont {Kraus},
  \citenamefont {Ringel},\ and\ \citenamefont {Zilberberg}}]{kraus2013four}%
  \BibitemOpen
  \bibfield  {author} {\bibinfo {author} {\bibfnamefont {Y.~E.}\ \bibnamefont
  {Kraus}}, \bibinfo {author} {\bibfnamefont {Z.}~\bibnamefont {Ringel}}, \
  and\ \bibinfo {author} {\bibfnamefont {O.}~\bibnamefont {Zilberberg}},\
  }\href@noop {} {\bibfield  {journal} {\bibinfo  {journal} {Physical review
  letters}\ }\textbf {\bibinfo {volume} {111}},\ \bibinfo {pages} {226401}
  (\bibinfo {year} {2013})}\BibitemShut {NoStop}%
\bibitem [{\citenamefont {Lohse}\ \emph {et~al.}(2018)\citenamefont {Lohse},
  \citenamefont {Schweizer}, \citenamefont {Price}, \citenamefont
  {Zilberberg},\ and\ \citenamefont {Bloch}}]{lohse2018exploring}%
  \BibitemOpen
  \bibfield  {author} {\bibinfo {author} {\bibfnamefont {M.}~\bibnamefont
  {Lohse}}, \bibinfo {author} {\bibfnamefont {C.}~\bibnamefont {Schweizer}},
  \bibinfo {author} {\bibfnamefont {H.~M.}\ \bibnamefont {Price}}, \bibinfo
  {author} {\bibfnamefont {O.}~\bibnamefont {Zilberberg}}, \ and\ \bibinfo
  {author} {\bibfnamefont {I.}~\bibnamefont {Bloch}},\ }\href@noop {}
  {\bibfield  {journal} {\bibinfo  {journal} {Nature}\ }\textbf {\bibinfo
  {volume} {553}},\ \bibinfo {pages} {55} (\bibinfo {year} {2018})}\BibitemShut
  {NoStop}%
\bibitem [{\citenamefont {Zilberberg}\ \emph {et~al.}(2018)\citenamefont
  {Zilberberg}, \citenamefont {Huang}, \citenamefont {Guglielmon},
  \citenamefont {Wang}, \citenamefont {Chen}, \citenamefont {Kraus},\ and\
  \citenamefont {Rechtsman}}]{zilberberg2018photonic}%
  \BibitemOpen
  \bibfield  {author} {\bibinfo {author} {\bibfnamefont {O.}~\bibnamefont
  {Zilberberg}}, \bibinfo {author} {\bibfnamefont {S.}~\bibnamefont {Huang}},
  \bibinfo {author} {\bibfnamefont {J.}~\bibnamefont {Guglielmon}}, \bibinfo
  {author} {\bibfnamefont {M.}~\bibnamefont {Wang}}, \bibinfo {author}
  {\bibfnamefont {K.~P.}\ \bibnamefont {Chen}}, \bibinfo {author}
  {\bibfnamefont {Y.~E.}\ \bibnamefont {Kraus}}, \ and\ \bibinfo {author}
  {\bibfnamefont {M.~C.}\ \bibnamefont {Rechtsman}},\ }\href@noop {} {\bibfield
   {journal} {\bibinfo  {journal} {Nature}\ }\textbf {\bibinfo {volume}
  {553}},\ \bibinfo {pages} {59} (\bibinfo {year} {2018})}\BibitemShut
  {NoStop}%
\bibitem [{\citenamefont {Qi}\ \emph {et~al.}(2008)\citenamefont {Qi},
  \citenamefont {Hughes},\ and\ \citenamefont {Zhang}}]{qi2008topological}%
  \BibitemOpen
  \bibfield  {author} {\bibinfo {author} {\bibfnamefont {X.-L.}\ \bibnamefont
  {Qi}}, \bibinfo {author} {\bibfnamefont {T.~L.}\ \bibnamefont {Hughes}}, \
  and\ \bibinfo {author} {\bibfnamefont {S.-C.}\ \bibnamefont {Zhang}},\
  }\href@noop {} {\bibfield  {journal} {\bibinfo  {journal} {Physical Review
  B}\ }\textbf {\bibinfo {volume} {78}},\ \bibinfo {pages} {195424} (\bibinfo
  {year} {2008})}\BibitemShut {NoStop}%
\bibitem [{\citenamefont {Hatsugai}\ and\ \citenamefont
  {Kohmoto}(1990)}]{hahaha}%
  \BibitemOpen
  \bibfield  {author} {\bibinfo {author} {\bibfnamefont {Y.}~\bibnamefont
  {Hatsugai}}\ and\ \bibinfo {author} {\bibfnamefont {M.}~\bibnamefont
  {Kohmoto}},\ }\href {\doibase 10.1103/PhysRevB.42.8282} {\bibfield  {journal}
  {\bibinfo  {journal} {Phys. Rev. B}\ }\textbf {\bibinfo {volume} {42}},\
  \bibinfo {pages} {8282} (\bibinfo {year} {1990})}\BibitemShut {NoStop}%
\bibitem [{\citenamefont {Lin}\ and\ \citenamefont
  {Hughes}(2017)}]{lin2017topological}%
  \BibitemOpen
  \bibfield  {author} {\bibinfo {author} {\bibfnamefont {M.}~\bibnamefont
  {Lin}}\ and\ \bibinfo {author} {\bibfnamefont {T.~L.}\ \bibnamefont
  {Hughes}},\ }\href@noop {} {\bibfield  {journal} {\bibinfo  {journal} {arXiv
  preprint arXiv:1708.08457}\ } (\bibinfo {year} {2017})}\BibitemShut {NoStop}%
\bibitem [{\citenamefont {Hashimoto}\ \emph {et~al.}(2017)\citenamefont
  {Hashimoto}, \citenamefont {Wu},\ and\ \citenamefont
  {Kimura}}]{hashimoto2017edge}%
  \BibitemOpen
  \bibfield  {author} {\bibinfo {author} {\bibfnamefont {K.}~\bibnamefont
  {Hashimoto}}, \bibinfo {author} {\bibfnamefont {X.}~\bibnamefont {Wu}}, \
  and\ \bibinfo {author} {\bibfnamefont {T.}~\bibnamefont {Kimura}},\
  }\href@noop {} {\bibfield  {journal} {\bibinfo  {journal} {Physical Review
  B}\ }\textbf {\bibinfo {volume} {95}},\ \bibinfo {pages} {165443} (\bibinfo
  {year} {2017})}\BibitemShut {NoStop}%
\bibitem [{\citenamefont {Langbehn}\ \emph {et~al.}(2017)\citenamefont
  {Langbehn}, \citenamefont {Peng}, \citenamefont {Trifunovic}, \citenamefont
  {von Oppen},\ and\ \citenamefont {Brouwer}}]{langbehn2017reflection}%
  \BibitemOpen
  \bibfield  {author} {\bibinfo {author} {\bibfnamefont {J.}~\bibnamefont
  {Langbehn}}, \bibinfo {author} {\bibfnamefont {Y.}~\bibnamefont {Peng}},
  \bibinfo {author} {\bibfnamefont {L.}~\bibnamefont {Trifunovic}}, \bibinfo
  {author} {\bibfnamefont {F.}~\bibnamefont {von Oppen}}, \ and\ \bibinfo
  {author} {\bibfnamefont {P.~W.}\ \bibnamefont {Brouwer}},\ }\href@noop {}
  {\bibfield  {journal} {\bibinfo  {journal} {Physical review letters}\
  }\textbf {\bibinfo {volume} {119}},\ \bibinfo {pages} {246401} (\bibinfo
  {year} {2017})}\BibitemShut {NoStop}%
\bibitem [{\citenamefont {Benalcazar}\ \emph
  {et~al.}(2017{\natexlab{a}})\citenamefont {Benalcazar}, \citenamefont
  {Bernevig},\ and\ \citenamefont {Hughes}}]{benalcazar2017quantized}%
  \BibitemOpen
  \bibfield  {author} {\bibinfo {author} {\bibfnamefont {W.~A.}\ \bibnamefont
  {Benalcazar}}, \bibinfo {author} {\bibfnamefont {B.~A.}\ \bibnamefont
  {Bernevig}}, \ and\ \bibinfo {author} {\bibfnamefont {T.~L.}\ \bibnamefont
  {Hughes}},\ }\href@noop {} {\bibfield  {journal} {\bibinfo  {journal}
  {Science}\ }\textbf {\bibinfo {volume} {357}},\ \bibinfo {pages} {61}
  (\bibinfo {year} {2017}{\natexlab{a}})}\BibitemShut {NoStop}%
\bibitem [{\citenamefont {Trifunovic}\ and\ \citenamefont
  {Brouwer}(2018)}]{trifunovic2018higher}%
  \BibitemOpen
  \bibfield  {author} {\bibinfo {author} {\bibfnamefont {L.}~\bibnamefont
  {Trifunovic}}\ and\ \bibinfo {author} {\bibfnamefont {P.}~\bibnamefont
  {Brouwer}},\ }\href@noop {} {\bibfield  {journal} {\bibinfo  {journal} {arXiv
  preprint arXiv:1805.02598}\ } (\bibinfo {year} {2018})}\BibitemShut {NoStop}%
\bibitem [{\citenamefont {Geier}\ \emph {et~al.}(2018)\citenamefont {Geier},
  \citenamefont {Trifunovic}, \citenamefont {Hoskam},\ and\ \citenamefont
  {Brouwer}}]{geier2018second}%
  \BibitemOpen
  \bibfield  {author} {\bibinfo {author} {\bibfnamefont {M.}~\bibnamefont
  {Geier}}, \bibinfo {author} {\bibfnamefont {L.}~\bibnamefont {Trifunovic}},
  \bibinfo {author} {\bibfnamefont {M.}~\bibnamefont {Hoskam}}, \ and\ \bibinfo
  {author} {\bibfnamefont {P.~W.}\ \bibnamefont {Brouwer}},\ }\href@noop {}
  {\bibfield  {journal} {\bibinfo  {journal} {Physical Review B}\ }\textbf
  {\bibinfo {volume} {97}},\ \bibinfo {pages} {205135} (\bibinfo {year}
  {2018})}\BibitemShut {NoStop}%
\bibitem [{\citenamefont {Schindler}\ \emph {et~al.}(2018)\citenamefont
  {Schindler}, \citenamefont {Wang}, \citenamefont {Vergniory}, \citenamefont
  {Cook}, \citenamefont {Murani}, \citenamefont {Sengupta}, \citenamefont
  {Kasumov}, \citenamefont {Deblock}, \citenamefont {Jeon}, \citenamefont
  {Drozdov} \emph {et~al.}}]{schindler2018higher}%
  \BibitemOpen
  \bibfield  {author} {\bibinfo {author} {\bibfnamefont {F.}~\bibnamefont
  {Schindler}}, \bibinfo {author} {\bibfnamefont {Z.}~\bibnamefont {Wang}},
  \bibinfo {author} {\bibfnamefont {M.~G.}\ \bibnamefont {Vergniory}}, \bibinfo
  {author} {\bibfnamefont {A.~M.}\ \bibnamefont {Cook}}, \bibinfo {author}
  {\bibfnamefont {A.}~\bibnamefont {Murani}}, \bibinfo {author} {\bibfnamefont
  {S.}~\bibnamefont {Sengupta}}, \bibinfo {author} {\bibfnamefont {A.~Y.}\
  \bibnamefont {Kasumov}}, \bibinfo {author} {\bibfnamefont {R.}~\bibnamefont
  {Deblock}}, \bibinfo {author} {\bibfnamefont {S.}~\bibnamefont {Jeon}},
  \bibinfo {author} {\bibfnamefont {I.}~\bibnamefont {Drozdov}},  \emph
  {et~al.},\ }\href@noop {} {\bibfield  {journal} {\bibinfo  {journal} {Nature
  Physics}\ ,\ \bibinfo {pages} {1}} (\bibinfo {year} {2018})}\BibitemShut
  {NoStop}%
\bibitem [{\citenamefont {Wang}\ \emph {et~al.}(2018)\citenamefont {Wang},
  \citenamefont {Wieder}, \citenamefont {Li}, \citenamefont {Yan},\ and\
  \citenamefont {Bernevig}}]{wang2018higher}%
  \BibitemOpen
  \bibfield  {author} {\bibinfo {author} {\bibfnamefont {Z.}~\bibnamefont
  {Wang}}, \bibinfo {author} {\bibfnamefont {B.~J.}\ \bibnamefont {Wieder}},
  \bibinfo {author} {\bibfnamefont {J.}~\bibnamefont {Li}}, \bibinfo {author}
  {\bibfnamefont {B.}~\bibnamefont {Yan}}, \ and\ \bibinfo {author}
  {\bibfnamefont {B.~A.}\ \bibnamefont {Bernevig}},\ }\href@noop {} {\bibfield
  {journal} {\bibinfo  {journal} {arXiv preprint arXiv:1806.11116}\ } (\bibinfo
  {year} {2018})}\BibitemShut {NoStop}%
\bibitem [{\citenamefont {Ezawa}(2018)}]{ezawa2018minimal}%
  \BibitemOpen
  \bibfield  {author} {\bibinfo {author} {\bibfnamefont {M.}~\bibnamefont
  {Ezawa}},\ }\href@noop {} {\bibfield  {journal} {\bibinfo  {journal} {arXiv
  preprint arXiv:1801.00437}\ } (\bibinfo {year} {2018})}\BibitemShut {NoStop}%
\bibitem [{\citenamefont {Serra-Garcia}\ \emph {et~al.}(2018)\citenamefont
  {Serra-Garcia}, \citenamefont {Peri}, \citenamefont {S{\"u}sstrunk},
  \citenamefont {Bilal}, \citenamefont {Larsen}, \citenamefont {Villanueva},\
  and\ \citenamefont {Huber}}]{serra2018observation}%
  \BibitemOpen
  \bibfield  {author} {\bibinfo {author} {\bibfnamefont {M.}~\bibnamefont
  {Serra-Garcia}}, \bibinfo {author} {\bibfnamefont {V.}~\bibnamefont {Peri}},
  \bibinfo {author} {\bibfnamefont {R.}~\bibnamefont {S{\"u}sstrunk}}, \bibinfo
  {author} {\bibfnamefont {O.~R.}\ \bibnamefont {Bilal}}, \bibinfo {author}
  {\bibfnamefont {T.}~\bibnamefont {Larsen}}, \bibinfo {author} {\bibfnamefont
  {L.~G.}\ \bibnamefont {Villanueva}}, \ and\ \bibinfo {author} {\bibfnamefont
  {S.~D.}\ \bibnamefont {Huber}},\ }\href@noop {} {\bibfield  {journal}
  {\bibinfo  {journal} {Nature}\ }\textbf {\bibinfo {volume} {555}},\ \bibinfo
  {pages} {342} (\bibinfo {year} {2018})}\BibitemShut {NoStop}%
\bibitem [{\citenamefont {Petrides}\ \emph {et~al.}(2018)\citenamefont
  {Petrides}, \citenamefont {Price},\ and\ \citenamefont
  {Zilberberg}}]{petrides2018six}%
  \BibitemOpen
  \bibfield  {author} {\bibinfo {author} {\bibfnamefont {I.}~\bibnamefont
  {Petrides}}, \bibinfo {author} {\bibfnamefont {H.~M.}\ \bibnamefont {Price}},
  \ and\ \bibinfo {author} {\bibfnamefont {O.}~\bibnamefont {Zilberberg}},\
  }\href@noop {} {\bibfield  {journal} {\bibinfo  {journal} {Physical Review
  B}\ }\textbf {\bibinfo {volume} {98}},\ \bibinfo {pages} {125431} (\bibinfo
  {year} {2018})}\BibitemShut {NoStop}%
\bibitem [{\citenamefont {Creutz}(1999)}]{creutz1999m}%
  \BibitemOpen
  \bibfield  {author} {\bibinfo {author} {\bibfnamefont {M.}~\bibnamefont
  {Creutz}},\ }\href@noop {} {\bibfield  {journal} {\bibinfo  {journal} {Phys.
  Rev. Lett.}\ }\textbf {\bibinfo {volume} {83}},\ \bibinfo {pages} {2636}
  (\bibinfo {year} {1999})}\BibitemShut {NoStop}%
\bibitem [{\citenamefont {Peierls}(1933)}]{Peierls:1933ZPhys}%
  \BibitemOpen
  \bibfield  {author} {\bibinfo {author} {\bibfnamefont {R.}~\bibnamefont
  {Peierls}},\ }\href@noop {} {\bibfield  {journal} {\bibinfo  {journal}
  {Zeitschrift f{\"u}r Physik}\ }\textbf {\bibinfo {volume} {80}},\ \bibinfo
  {pages} {763} (\bibinfo {year} {1933})}\BibitemShut {NoStop}%
\bibitem [{Note1()}]{Note1}%
  \BibitemOpen
  \bibinfo {note} {For convenience we drop the dependence on the remaining
  coordinates $\protect \text {m}_\sigma $ and $\protect \text {k}_\rho $ since
  the Hamiltonian $\protect \mathaccentV {hat}05E{h}_{\mu }(\protect \text
  {k}_\nu )$ is independent of them}\BibitemShut {NoStop}%
\bibitem [{\citenamefont {Su}\ \emph {et~al.}(1979)\citenamefont {Su},
  \citenamefont {Schrieffer},\ and\ \citenamefont {Heeger}}]{su1979solitons}%
  \BibitemOpen
  \bibfield  {author} {\bibinfo {author} {\bibfnamefont {W.}~\bibnamefont
  {Su}}, \bibinfo {author} {\bibfnamefont {J.}~\bibnamefont {Schrieffer}}, \
  and\ \bibinfo {author} {\bibfnamefont {A.~J.}\ \bibnamefont {Heeger}},\
  }\href@noop {} {\bibfield  {journal} {\bibinfo  {journal} {Physical review
  letters}\ }\textbf {\bibinfo {volume} {42}},\ \bibinfo {pages} {1698}
  (\bibinfo {year} {1979})}\BibitemShut {NoStop}%
\bibitem [{\citenamefont {King-Smith}\ and\ \citenamefont
  {Vanderbilt}(1993)}]{king1993theory}%
  \BibitemOpen
  \bibfield  {author} {\bibinfo {author} {\bibfnamefont {R.}~\bibnamefont
  {King-Smith}}\ and\ \bibinfo {author} {\bibfnamefont {D.}~\bibnamefont
  {Vanderbilt}},\ }\href@noop {} {\bibfield  {journal} {\bibinfo  {journal}
  {Physical Review B}\ }\textbf {\bibinfo {volume} {47}},\ \bibinfo {pages}
  {1651} (\bibinfo {year} {1993})}\BibitemShut {NoStop}%
\bibitem [{\citenamefont {Vanderbilt}\ and\ \citenamefont
  {King-Smith}(1993)}]{vanderbilt1993electric}%
  \BibitemOpen
  \bibfield  {author} {\bibinfo {author} {\bibfnamefont {D.}~\bibnamefont
  {Vanderbilt}}\ and\ \bibinfo {author} {\bibfnamefont {R.}~\bibnamefont
  {King-Smith}},\ }\href@noop {} {\bibfield  {journal} {\bibinfo  {journal}
  {Physical Review B}\ }\textbf {\bibinfo {volume} {48}},\ \bibinfo {pages}
  {4442} (\bibinfo {year} {1993})}\BibitemShut {NoStop}%
\bibitem [{\citenamefont {Jackiw}\ and\ \citenamefont
  {Rebbi}(1976)}]{jackiw1976solitons}%
  \BibitemOpen
  \bibfield  {author} {\bibinfo {author} {\bibfnamefont {R.}~\bibnamefont
  {Jackiw}}\ and\ \bibinfo {author} {\bibfnamefont {C.}~\bibnamefont {Rebbi}},\
  }\href@noop {} {\bibfield  {journal} {\bibinfo  {journal} {Physical Review
  D}\ }\textbf {\bibinfo {volume} {13}},\ \bibinfo {pages} {3398} (\bibinfo
  {year} {1976})}\BibitemShut {NoStop}%
\bibitem [{\citenamefont {Graf}\ and\ \citenamefont
  {Shapiro}(2018)}]{graf2018bulk}%
  \BibitemOpen
  \bibfield  {author} {\bibinfo {author} {\bibfnamefont {G.~M.}\ \bibnamefont
  {Graf}}\ and\ \bibinfo {author} {\bibfnamefont {J.}~\bibnamefont {Shapiro}},\
  }\href@noop {} {\bibfield  {journal} {\bibinfo  {journal} {Communications in
  Mathematical Physics}\ }\textbf {\bibinfo {volume} {363}},\ \bibinfo {pages}
  {829} (\bibinfo {year} {2018})}\BibitemShut {NoStop}%
\bibitem [{exa()}]{example}%
  \BibitemOpen
  \href@noop {} {}\bibinfo {note} {Taking $t_z > 0$ and $t_{xz} = 0$ in the 2D
  ancestor Hamiltonian $\hat{H}_{\mu\nu}(\tilde{\mathbf{m}},\tilde{\mathbf{k}}
  )$ leads, upon dimensional reduction, to a topologically trivial model.
  However, localized states appear at its interface with an insulator with $t_z
  < 0$ and $t_{xz} = 0$ due to a combination nonzero surface and bulk dipole
  moments.}\BibitemShut {Stop}%
\bibitem [{\citenamefont {Benalcazar}\ \emph
  {et~al.}(2017{\natexlab{b}})\citenamefont {Benalcazar}, \citenamefont
  {Bernevig},\ and\ \citenamefont {Hughes}}]{benalcazar2017electric}%
  \BibitemOpen
  \bibfield  {author} {\bibinfo {author} {\bibfnamefont {W.~A.}\ \bibnamefont
  {Benalcazar}}, \bibinfo {author} {\bibfnamefont {B.~A.}\ \bibnamefont
  {Bernevig}}, \ and\ \bibinfo {author} {\bibfnamefont {T.~L.}\ \bibnamefont
  {Hughes}},\ }\href@noop {} {\bibfield  {journal} {\bibinfo  {journal}
  {Physical Review B}\ }\textbf {\bibinfo {volume} {96}},\ \bibinfo {pages}
  {245115} (\bibinfo {year} {2017}{\natexlab{b}})}\BibitemShut {NoStop}%
\bibitem [{app()}]{appendix}%
  \BibitemOpen
  \href@noop {} {}\bibinfo {note} {Supplementary material: We derive the charge
  accumulation at the corner of 2D materials, show the multipole description of
  macroscopic material and calculate the 2$^{\mathrm{nd}}$ Chern forms of the
  three examples considered in the main text.}\BibitemShut {Stop}%
\end{thebibliography}%

\appendix

\newpage
\cleardoublepage
\renewcommand{\figurename}{Supplementary Material Figure}

\onecolumngrid
\begin{center}
	\textbf{\normalsize Supplemental Material for}\\
	\vspace{3mm}
	\textbf{\large Quantized corner charges, topological pumps and the \Snd Chern number}
	\vspace{4mm}
	
	{ Ioannis\ Petrides, and Oded Zilberberg}\\
	\vspace{1mm}
	\textit{\small Institute for Theoretical Physics, ETH Z\"urich, 8093 Z\"urich, Switzerland\\
	}
	
	\vspace{5mm}
\end{center}
%%%%%%%%%% Merge with supplemental materials %%%%%%%%%%
%%%%%%%%%% Prefix a "S" to all equations, figures, tables and reset the counter %%%%%%%%%%
\setcounter{equation}{0}
\setcounter{section}{0}
\setcounter{figure}{0}
\setcounter{table}{0}
\setcounter{page}{1}
\makeatletter
\renewcommand{\bibnumfmt}[1]{[#1]}
\renewcommand{\citenumfont}[1]{#1}
%%%%%%%%%% Prefix a "S" to all equations, figures, tables and reset the counter %%%%%%%%%%

%\setcounter{enumi}{1}
%\renewcommand{\theequation}{\Roman{enumi}.\arabic{equation}}

\section{Charge accumulation at the corner of 2D materials}
\label{sec:charge accumulation}
In this Section, we calculate the charge accumulation at the corner of a 2D macroscopic material described by the three models discussed in the main text and show its relation to the \Snd Chern flux. 
Specifically, we find the low-energy continuoum description of the momentum space Hamiltonian and solve for localised states at the corner.
We calculate the charge accumulation at the corner by constructing a smooth boundary with the vacuum and perturbatively expanding the variation of the Hamiltonian up to second order. 
Finally, we show the quantization of the accumulated charge in the limit where chiral symmetry is restored.

\subsection{Model I}\label{subsec:Model I}
We consider the Hamiltonian 
$\hat{h}^{\pi}_{xy}(\tilde{\mathbf{k}}) = \sum_{\tilde{\mathbf{m}}}\left[\hat{H}_{xz}(\tilde{\mathbf{m}},\tilde{\mathbf{k}} )+\hat{H}_{yw}(\tilde{\mathbf{m}},\tilde{\mathbf{k}}) + \Delta\hat{H}^{\pi} _{xy} (\tilde{\mathbf{m}},\tilde{\mathbf{k}})\right]$, 
defined in the main text, where $\tilde{\mathbf{k}}$ is treated as an external two-dimensional parameter space. 
The resulting lattice is made out of SSH chains  [see Eq.~\eqref{eq:Creutz} in the main text] in both the $x$- and $y$-directions, and each $xy$-plaquette is threaded by a magnetic field with $\pi$ flux quanta, see Fig.~2(a) in the main text. 
As a function of $\tilde{\mathbf{k}}$, the spectrum of $\hat{h}^{\pi}_{xy}(\tilde{\mathbf{k}})$ has: (i) 1D edge states appearing below zero energy that merge into the bulk, and (ii) zero-energy 0D corner states that merge into the edge or bulk spectrum at the gap closing points, see Fig.~2(c).

Linearising the dynamics of $\hat{h}^{\pi}_{xy}(\tilde{\mathbf{k}})$ around zero energy we find that the low-energy theory is given by a single Dirac cone around $\textbf{k} \sim (\pi,\pi)$, described by the Hamiltonian 
\begin{eqnarray}
\hat{h}^{\pi}_{xy}(\tilde{\mathbf{k}}) = \int\limits_{|\mathbf{{k}}|<\Lambda} \mathrm{d}^2 \mathbf{{k}} \,\mathbf{{d}}_0\cdot\mathbf{\Gamma}\,,
\label{eq:Model I}
\end{eqnarray}
where $\mathbf{d}_0=\{\mu_0 (\tilde{\mathbf{k}}), \mu_1 (\tilde{\mathbf{k}}), \mu_2 (\tilde{\mathbf{k}}), {v}_x (\tilde{\mathbf{k}}) \mathrm{k}_x, {v}_y (\tilde{\mathbf{k}}) \mathrm{k}_y  \}$, $v_\mu (\mathrm{k}_\nu)= J_{\mu\nu}^+  $, $\mu_2 (\mathrm{k}_z)= J_{xz}^+ -J_{xz}^- $, $\mu_1 (\mathrm{k}_w)= J_{yw}^+ -J_{yw}^- $, $\mu_0$ is a chiral-breaking mass (in general can be $\tilde{\mathbf{k}}$ depended), $\Lambda$ is the cut-off energy scale of the low-energy theory above which the approximation is not valid, and $\mathbf{\Gamma} = \{ \Gamma_0, \Gamma_1 ,\Gamma_2 ,\Gamma_3 , \Gamma_4 \}$ are five matrices,
\begin{align}	
	\Gamma_0 =\begin{pmatrix}     
		-\mathds{1} & 0 \\
		0 & \mathds{1} 
	\end{pmatrix}\,,
	\Gamma_1 =\begin{pmatrix}  
		0 &  \sigma_z \\
		\sigma_z & 0 
	\end{pmatrix}\,,
	\Gamma_2 =\begin{pmatrix}    
		0 & \sigma_x\\
		\sigma_x & 0 
	\end{pmatrix}\,,
	\Gamma_3 =\begin{pmatrix}    
		0 &- i\mathds{1} \\
		i\mathds{1} & 0  
	\end{pmatrix}\,,
	\Gamma_4 =\begin{pmatrix}     
		0 & \sigma_y\\
		\sigma_y & 0   
	\end{pmatrix}\,, 
\label{eq:basis I}
\end{align}    
satisfying the Clifford algebra, $\{\Gamma_i , \Gamma_j\} = 2\delta_{ij}\mathds{1}$, with $\{\sigma_x,\sigma_y,\sigma_z\}$ the Pauli matrices.  
The above model is equivalent to a free two-dimensional Dirac particle with linear dispersion $v_x$ ($v_y$) in the $x$- ($y$-) direction, in a background potential $\bm{\mu} = \{\mu_0,\mu_1,\mu_2\}$. 
The localised states around zero energy [cf.~Fig.~(2)(c) and (d)] are found by solving the continuoum Hamiltonian $\mathbf{{d}}_c\cdot\mathbf{\Gamma}$, where ${\textbf{d}}_c = \{\mu_0 , \mu_1 , \mu_2 , -{v}_x i\partial_x, -{v}_y  i\partial_y \}$ (we drop the $\tilde{\mathbf{k}}$-dependence for simplicity), with a plane wave ansatz $\psi = e^{i (\text{k}_x x + \text{k}_y y)}\phi$, where $\phi$ are the Bloch eigenvectors of ${\textbf{d}}_0 \cdot\bm{\Gamma}$. 
Since we are interested in finding corner states, i.e., states that are localised in both dimensions, we place our system on a semi-infinite space in $\mathbb{R}^2 _{\ge 0}$, where $x\ge0$ and $y\ge0$ (with Dirichlet boundary condition for $x<0$ and $y<0$). 
Assuming $\mu_0$ can be treated perturbatively, we expand up to first order to find a single normalisable solution below zero energy (assuming $\frac{\mu_1}{v_x} > 0$ and $\frac{\mu_2}{v_y} >0$)
\begin{align}
\psi =e^{-\int\limits_0 ^x \frac{m_1}{v_x} \text{d}x'}e^{-\int\limits_0 ^y \frac{m_2}{v_y} \text{d}y'}\begin{pmatrix}     
1 \\
0 \\
0 \\
0
\end{pmatrix}\,,
\label{eq:localised solution}
\end{align}
with negative eigenvalue $- \mu_0$.
The above wavefunction is localised in both dimensions around the origin of the semi-infinite space and decays exponentially in space.

In order to calculate the charge accumulation at the interface between the macroscopic material described by $\hat{h}^{\pi}_{xy}(\tilde{\mathbf{k}})$ and the vacuum, we extend the semi-infinite space $\mathbb{R}^2 _{\ge 0}$ to $\mathbb{R}^2$ and introduce domain walls in the $\mu_1 (x)$ and $\mu_2 (y)$ parameters.
Namely, the domain walls interpolates between: (i) a ground state with asymptotic value $\mu_1 (x\to \infty) = \mu_1(\tilde{\mathbf{k}})$ and a vacuum state with $\mu_1 (x\to -\infty) < 0$, and (ii) a ground state with asymptotic value $\mu_2 (y\to \infty) = \mu_2(\tilde{\mathbf{k}})$ and a vacuum state with $\mu_2 (y\to -\infty) < 0$ [see Fig.~2(a)]. 
This results in having the desired Hamiltonian $\hat{h}_{xy}(\tilde{\mathbf{k}})$ in the upper-right quadrant of $\mathbb{R}^2$, continuously connected to the vacuum state in the lower-right (upper-left) quadrant with no $y$-($x$-) localised solutions. 
In general, any arbitrary, continuous interpolation between $\mu_1 (x\to \infty)$ and $\mu_1 (x\to- \infty)$ [and similarly for $\mu_2 (y)$] is sufficient, however, a corner state will manifest only at the intersection of two domain walls.

To calculate the charge density and charge accumulation at the corners of the 2D model with the domain walls defined above, cf.~Fig~2(a), we first define the corresponding (2+1)D Lagrangian
\begin{align}
	\mathcal{L} =\bar{\psi} \left(i\slashed{\partial} - \bm{\mu}\cdot \bm{\tau}\right)\psi \,,
\end{align}
where $\slashed{\partial} = v_\mu \gamma^\mu \partial_\mu$, with $\mu = \{0,1,2\}$, is a sum of derivatives multiplied by the matrices $\gamma_0 = \Gamma_0$, $\gamma_1 = \Gamma_0 \Gamma_1$, $\gamma_2 = \Gamma_0 \Gamma_2$, $\bm{\mu} = \{\mu_0,\mu_1,\mu_2\}$ is a real-valued vector describing a background potential on the basis $\bm{\tau} = \{\mathds{1}, \Gamma_0 \Gamma_3 , \Gamma_0 \Gamma_4\}$, and $\bar{\psi}=\psi^\dagger \gamma_0$ is the ``anti-matter" field. 
The above Lagrangian describes a free massless two-dimensional Dirac field in a background ``Yukawa" potential $\bm{\mu}\cdot\bm{\tau}$. 
Importantly, the equations of motion derived from $\mathcal{L}$ are the same as those obtained by the continuoum Hamiltonian $i\partial_0 \psi = \bm{\textbf{d}}_c\cdot\bm{\Gamma} \psi$. 
Coupling the theory to a $U(1)$ gauge field (i.e., to electromagnetism), the many-body ground state expectation value of the conserved 3-current at position $\bar{x} = \left(t,x,y\right)$ is given by 
\begin{align}
	\big\langle\hat{j}^\mu (\bar{x})\big\rangle =-\frac{i}{2}\text{tr}\left(\gamma^\mu S^r (\bar{x},\bar{x}) + \gamma^\mu S^l (\bar{x},\bar{x})\right) \,,
	\label{eq:3-current}
\end{align}
where $S^r (\bar{x},\bar{y})$ and $S^l (\bar{x},\bar{y})$ are the Green's functions associated with the equations of motion for the fields $\psi$ and $\bar{\psi}$
\begin{align}
	 \left(i\vec{\slashed{\partial}}_{\bar{x}} - \bm{\mu}\cdot \bm{\tau}\right) S^r (\bar{x},\bar{y}) =\mathds{1} \delta(\bar{x}-\bar{y}) \,,
	 \hspace{10pt}\text{and}\hspace{10pt}
	   S^l (\bar{x},\bar{y})\left(i\cev{\slashed{\partial}}_{\bar{y}} -\bm{\mu}\cdot \bm{\tau}\right) =\mathds{1} \delta(\bar{x}-\bar{y}) \,.
\end{align}
Here, $\cev{\slashed{\partial}}_{\bar{y}}$ ($\vec{\slashed{\partial}}_{\bar{x}}$) is the derivative acting to the left (right) with respect to the $\bar{y}$ ($\bar{x}$) coordinates. 
Using the Fourier transforms:
\begin{align}
	\delta(\bar{x}-\bar{y}) &=\int \frac{\mathrm{d}^3 {\bar{k}}}{(2\pi)^3}e^{-i\bar{k}\cdot(\bar{x}-\bar{y})}\,,\\
	S^l (\bar{x},\bar{y})&=\int \frac{\mathrm{d}^3 {\bar{k}}}{(2\pi)^3}e^{-i\bar{k}\cdot(\bar{x}-\bar{y})}S^l (\bar{x},\bar{k})\,,\\
	S^r (\bar{x},\bar{y})&=\int \frac{\mathrm{d}^3 {\bar{k}}}{(2\pi)^3}e^{-i\bar{k}\cdot(\bar{x}-\bar{y})}S^r (\bar{x},\bar{k})\,,
\end{align}
where $\bar{k}=\{\omega,\mathrm{{k}}_x,\mathrm{{k}}_y\}$ is the 3-momentum, the Green's functions can be written as
\begin{align}
	S^r (\bar{x},\bar{k})&=\frac{1}{\slashed{k}+i\vec{\slashed{\partial}}_{\bar{x}}-\bm{\mu}\cdot \bm{\tau}}\,,\\
	S^l (\bar{x},\bar{k})&=\frac{1}{-\slashed{k}+i\cev{\slashed{\partial}}_{\bar{x}}-\bm{\mu}\cdot \bm{\tau}}\,.
\end{align}

The charge accumulation in a closed region $C$ [e.g., see Fig.~2(a)] defined anywhere in $\mathbb{R}^2$ (i.e., anywhere in the material) is given by the integral of the zeroth component of the current $\big\langle\hat{j}^0 (\bar{x})\big\rangle $ 
\begin{align}
	q_C = \int_{C}	\big\langle\hat{j}^0 (\bar{x})\big\rangle \mathrm{d}^2 \mathbf{r}\,,
	\label{eq:charge accumulation}
\end{align}
where $\mathbf{r} = \{x,y\}$ are the spatial dimensions. 
In order to find an analytic expression for $q_C $, we assume that the domain wall interpolation is done smoothly and slowly enough, such that the Green's functions $S^r $ and $S^l $ can be perturbatively expanded in powers of the gradients
\begin{align}
S^r (\bar{x},\bar{k})\simeq	&\frac{1}{\slashed{k}-\bm{\mu}\cdot\bm{\tau}}
	+\frac{1}{\slashed{k}-\bm{\mu}\cdot\bm{\tau}}\cdot
	\left(-i\vec{\slashed{\partial}}_{\bar{x}}\right)
	\cdot
	\frac{1}{\slashed{k}-\bm{\mu}\cdot\bm{\tau}}
	+\frac{1}{\slashed{k}-\bm{\mu}\cdot\bm{\tau}}\cdot
	\left(-i\vec{\slashed{\partial}}_{\bar{x}}\right)
	\cdot
	\frac{1}{\slashed{k}-\bm{\mu}\cdot\bm{\tau}}\left(-i\vec{\slashed{\partial}}_{\bar{x}}\right)
	\cdot
	\frac{1}{\slashed{k}-\bm{\mu}\cdot\bm{\tau}}\,,\\
	S^l (\bar{x},\bar{k})\simeq	&\frac{1}{-\slashed{k}-\bm{\mu}\cdot\bm{\tau}}
	+\frac{1}{-\slashed{k}-\bm{\mu}\cdot\bm{\tau}}\cdot
	\left(-i\cev{\slashed{\partial}}_{\bar{x}}\right)
	\cdot
	\frac{1}{-\slashed{k}-\bm{\mu}\cdot\bm{\tau}}
	+\frac{1}{-\slashed{k}-\bm{\mu}\cdot\bm{\tau}}\cdot
	\left(-i\cev{\slashed{\partial}}_{\bar{x}}\right)
	\cdot
	\frac{1}{-\slashed{k}-\bm{\mu}\cdot\bm{\tau}}\left(-i\cev{\slashed{\partial}}_{\bar{x}}\right)
	\cdot
	\frac{1}{-\slashed{k}-\bm{\mu}\cdot\bm{\tau}}\,,
\end{align}
where we keep terms up to second order.
Substituting the above expansions into Eq.~\eqref{eq:charge accumulation}, we find that the first nonvanishing contributions are at second order 
\begin{align}
	q_C =\frac{3}{8\pi^2}\int_{\tilde{C}}\hat{\mathbf{\textbf{d}}}_0\cdot (\partial_{\mathrm{k}_x}\hat{\textbf{\text{d}}}_0 \times \partial_{\mathrm{k}_y}\hat{\mathbf{\textbf{d}}}_0\times \partial_{x}\hat{\mathbf{\textbf{d}}}_0 \times \partial_{y}\hat{\mathbf{\textbf{d}}}_0)\mathrm{d}^2 \mathbf{{k}}\mathrm{d}^2 \mathbf{r}\,,
	\label{eq:corner charge}
\end{align}
where $\hat{\mathbf{d}}_0=\frac{\mathbf{d}_0}{|\mathbf{d}_0|}$ is the normalised vector of the low-energy momentum-space Hamiltonian $\mathbf{{d}}_0\cdot\mathbf{\Gamma}$, $\tilde{C} = C\times [-\Lambda,\Lambda ]^2$ is the integration domain and $\Lambda$ is the cut-off energy scale of the low-energy theory obtained from the lattice model $\mathbf{{d}}\cdot\mathbf{\Gamma}$. 
Since the spatial dependence of $\hat{h}^{\pi}_{xy}(\tilde{\mathbf{k}}) $ is through the $\tilde{\mathbf{k}}$ parameters [i.e., $\tilde{\mathbf{k}}$ is a function of $\mathbf{r}$], we can change variables of integration in Eq.~\eqref{eq:corner charge} to obtain
\begin{align}
	q_C =\frac{3}{8\pi^2}\int_{\tilde{C}}\hat{\mathbf{\textbf{d}}}_0\cdot (\partial_{\mathrm{k}_x}\hat{\textbf{\text{d}}}_0 \times \partial_{\mathrm{k}_y}\hat{\mathbf{\textbf{d}}}_0\times \partial_{\mathrm{k}_z}\hat{\mathbf{\textbf{d}}}_0 \times \partial_{\mathrm{k}_w}\hat{\mathbf{\textbf{d}}}_0)\mathrm{d}^2 \mathbf{{k}}\mathrm{d}^2 \tilde{\mathbf{k}}\,,
	\label{eq:corner is 2nd CN}
\end{align}
where the integration is done over the region defined by $C$ in the $\tilde{\mathbf{k}}$-parameter space. 
This is exactly the expression of the \Snd Chern flux attached to a region in the 4D BZ of the ancestor model [cf., Eq.~\eqref{eq:2nd Chern Flux}].

In the case of a domain wall defined by the asymptotic values $\mu_i ( \infty) = \mu_i (\tilde{\mathbf{k}})$ and $\mu_i (-\infty) < 0$, with $i=1$ or $2$ [cf., Fig.~2(a)], we find that the charge accumulation in the limit $\mu_0\to 0$ (i.e, when chiral symmetry is restored) takes two values
\begin{align}
	\lim_{\mu_0\to0} |q_C | =&
		\frac{1}{2} \text{   or   }		0\,,
\end{align}
depending on the value of $\tilde{\mathbf{k}}$. 
In the nontrivial case $|q_C| = \frac{1}{2}$, the integration region defined by $\tilde{C}$ encloses a 4D Dirac point with a singular \Snd Chern number flux equal to $\pm\frac{1}{2}$.

\subsection{Model II}
We consider the Hamiltonian 
$\hat{h}^{0}_{xy}(\tilde{\mathbf{k}}) = \sum_{\tilde{\mathbf{m}}}\left[\hat{H}_{xz}(\tilde{\mathbf{m}},\tilde{\mathbf{k}} )+\hat{H}_{yw}(\tilde{\mathbf{m}},\tilde{\mathbf{k}}) + \Delta\hat{H}^{0} _{xy} (\tilde{\mathbf{m}},\tilde{\mathbf{k}})\right]$, 
defined in the main text, where $\tilde{\mathbf{k}}$ is treated as an external two-dimensional parameter space. 
The resulting lattice is made out of SSH chains $\hat{h}_{x}(\text{k}_z)$ [see Eq.~\eqref{eq:Creutz}] along the $x$-direction coupled to chains $\hat{h}_{y}(\text{k}_w )$ in the $y$-direction, see Fig.~3(a) for the real space lattice. 
The two components $\hat{h}_{x}(\text{k}_z)$ and $\hat{h}_{y}(\text{k}_w)$ are defined to have a relative hopping strength $t_x / t_y =\epsilon\ne 1$, since only in this case the spectrum has a finite gap at zero energy with corresponding 1D edge and 0D corner states, see Fig.~3(c).

The low-energy theory of $\hat{h}^{0}_{xy}(\tilde{\mathbf{k}})$ is given by a single conical spectrum around $\textbf{k} \sim (\pi,\pi)$, described by the Hamiltonian 
\begin{eqnarray}
\hat{h}^{\pi}_{xy}(\tilde{\mathbf{k}}) = \int \mathrm{d}^2 \mathbf{{k}} \,\mathbf{{d}}_0\cdot\mathbf{\Gamma}\,,
\label{eq:Model II}
\end{eqnarray}
where $\mathbf{d}_0=\{\mu_0 (\tilde{\mathbf{k}}), \mu_1 (\tilde{\mathbf{k}}), \epsilon\mu_2 (\tilde{\mathbf{k}}), {v}_x (\tilde{\mathbf{k}}) \mathrm{k}_x, \epsilon{v}_y (\tilde{\mathbf{k}}) \mathrm{k}_y  \}$, $v_\mu (\mathrm{k}_\nu)= J_{\mu\nu}^+  $, $\mu_2 (\mathrm{k}_z)= J_{xz}^+ -J_{xz}^- $, $\mu_1 (\mathrm{k}_w)= J_{yw}^+ -J_{yw}^- $, $\mu_0$ is a chiral-breaking mass (in general can be $\tilde{\mathbf{k}}$ depended), and $\mathbf{\Gamma} = \{ \Gamma_0, \Gamma_1 ,\Gamma_2 ,\Gamma_3 , \Gamma_4 \}$ are five matrices,
\begin{align}	
	\Gamma_0 =\begin{pmatrix}     
		-\mathds{1} & 0 \\
		0 & \mathds{1} 
	\end{pmatrix}\,,
	\Gamma_1 =\begin{pmatrix}  
		0 &  \mathds{1} \\
		\mathds{1} & 0 
	\end{pmatrix}\,,
	\Gamma_2 =\begin{pmatrix}    
		0 & \sigma_x\\
		\sigma_x & 0 
	\end{pmatrix}\,,
	\Gamma_3 =\begin{pmatrix}    
		0 &- i\sigma_z \\
		i\sigma_z & 0  
	\end{pmatrix}\,,
	\Gamma_4 =\begin{pmatrix}     
		0 & \sigma_y\\
		\sigma_y & 0   
	\end{pmatrix}\,, 
\label{eq:basis II}
\end{align}    
satisfying the chiral algebra, $\{\Gamma_0 , \Gamma_j\} = 2\delta_{0j}\mathds{1}$. 
The above model is equivalent to a free two-dimensional fermionic particle with linear dispersion $v_x$ ($v_y$) in the $x$- ($y$-) direction, in a background potential. 
As in Sec.~\ref{subsec:Model I}, the localised states around zero energy [cf.~Fig.~(3)(c)] are found by solving the continuoum Hamiltonian $\mathbf{{d}}_c\cdot\mathbf{\Gamma}$, where ${\textbf{d}}_c = \{\mu_0 , \mu_1 ,\epsilon \mu_2 , -{v}_x i\partial_x, -\epsilon{v}_y  i\partial_y \}$, with a plane wave ansatz $\psi = e^{i (\text{k}_x x + \text{k}_y y)}\phi$, where $\phi$ are the Bloch eigenvectors of ${\textbf{d}}_0 \cdot\bm{\Gamma}$. 
We place our system on a semi-infinite space in $\mathbb{R}^2 _{\ge 0}$, where $x\ge0$ and $y\ge0$ (with Dirichlet boundary condition for $x<0$ and $y<0$) and take the limit $\mu_0\to 0$. 
We expand up to first order and find a single normalisable solution below zero energy (assuming $\frac{\mu_1}{v_x} > 0$ and $\frac{\mu_2}{v_y} >0$)
\begin{align}
	\psi =e^{-\int\limits_0 ^x \frac{m_1}{v_x} \text{d}x'}e^{-\int\limits_0 ^y \frac{m_2}{v_y} \text{d}y'}\begin{pmatrix}     
		1 \\
		0 \\
		0 \\
		0
	\end{pmatrix}\,,
	\label{eq:localised solution II}
\end{align}
with negative eigenvalue $- \mu_0$.
The above wavefunction is localised in both dimensions around the origin of the semi-infinite space and decays exponentially in space.

Following a similar procedure as in Sec.~\ref{subsec:Model I}, we construct the interface between the macroscopic material described by $\hat{h}^{0}_{xy}(\tilde{\mathbf{k}})$ and the vacuum by extending the semi-infinite space $\mathbb{R}^2 _{\ge 0}$ to $\mathbb{R}^2$ and introducing domain walls in the $\mu_1 (x)$ and $\mu_2 (y)$ parameters.
The corresponding (2+1)D Lagrangian is given by
\begin{align}
	\mathcal{L} =\bar{\psi} \left(i\slashed{\partial} - \bm{\mu}\cdot \bm{\tau}\right)\psi \,,
\end{align}
where we adopt the same notation as in Sec.~\ref{subsec:Model I}.
We couple the theory to a $U(1)$ gauge field (i.e., to electromagnetism), and calculate Eq.~\eqref{eq:charge accumulation} up to second order in the gradients of $\mu_1$ and $\mu_2$.
We find that the first nonvanishing contribution to the charge accumulation in a closed region $C$ [eg., see Fig.~2(a)] is given by
\begin{align}
	q_C =\frac{3i \epsilon^2}{\pi^3}\int_{\tilde{C}}\int
	\frac{\omega^2 +\text{k}_x ^2 + \mu_1 ^2 - \mu_0 ^2}{\left(-\omega^2 +\text{k}_x ^2 + \mu_1 ^2 + \mu_0\right)^4} \mu_0\partial_{x} \mu_1 \partial_{y} \mu_2\,\mathrm{d}\omega
	\mathrm{d}^2 \mathbf{{k}}\mathrm{d}^2 \mathbf{r}\,,
	\label{eq:corner charge II}
\end{align}
where we have expanded around $\epsilon\sim 0$ and kept up to second order. 
Changing variables of integration $\mathbf{r}\to \tilde{\mathbf{k}}$ in Eq.~\eqref{eq:corner charge II}, we obtain an expression proportional to the \Snd Chern flux attached to a region in the 4D BZ of the ancestor model [cf., Eq.~\eqref{eq:2nd Chern Flux II}].
In the limit $\mu_0\to 0$, i.e., when chiral symmetry is restored, the charge accumulation becomes quantized to
\begin{align}
	\lim_{\mu_0\to0} |q_C | =&
	\frac{1}{2} \text{   or   }		0\,.
\end{align}
In the nontrivial case $|q_C| = \frac{1}{2}$ the integration region defined by $\tilde{C}$ encloses a 4D crossing point in the $(\mathrm{k}_x , \mathrm{k}_y , \mathrm{k}_z ,\mathrm{k}_w)$-parameter space.

\subsection{Model III}
We consider the Hamiltonian 
$\hat{h}^{0}_{xy}(\tilde{\mathbf{k}}) = \sum_{\tilde{\mathbf{m}}}\left[\hat{H}_{xz}(\tilde{\mathbf{m}},\tilde{\mathbf{k}} )+\hat{H}_{yw}(\tilde{\mathbf{m}},\tilde{\mathbf{k}}) + \Delta\hat{H}^{0} _{xy} (\tilde{\mathbf{m}},\tilde{\mathbf{k}})\right]$, 
defined in the main text, with an additional $\pi$-flux through the $xw$-plane.
The resulting lattice is made out of SSH chains $\hat{h}_{x}(\text{k}_z)$ along the $x$-direction coupled to alternating SSH chains $\hat{h}_{y}(\text{k}_w +\pi x)$ in the $y$-direction, see Fig.~3(d) in the main text. 
The two components $\hat{h}_{x}(\text{k}_z)$ and $\hat{h}_{y}(\text{k}_w)$ are defined to have a relative hopping strength $t_x / t_y =\epsilon\ne 1$, since in this case, the spectrum has a finite gap at zero energy with corresponding 1D edge and 0D corner states, see Fig.~3(f).

% [or $\tilde{\textbf{k}}^- \sim (0,\pi)$]

The low-energy theory of the model around $\textbf{k} \sim (\pi,\pi)$ and $\tilde{\textbf{k}}^+ \sim (0,0)$, is described by the Hamiltonian 
\begin{eqnarray}
\int \mathrm{d}^2 \mathbf{{k}} \,\mathbf{{d}}_0\cdot\mathbf{\Gamma}\,,
\label{eq:Model III}
\end{eqnarray}
where $\mathbf{d}_0=\{\mu_0 (\tilde{\mathbf{k}}), \mu_1 (\tilde{\mathbf{k}}), \epsilon\mu_2 (\tilde{\mathbf{k}}), {v} _x (\tilde{\mathbf{k}}) \mathrm{k}_x, \epsilon{v} _y (\tilde{\mathbf{k}}) \mathrm{k}_y  \}$, $v_\mu  = J_{\mu\nu}^+  $, $\mu_2 (\mathrm{k}_z)= J_{xz}^+ -J_{xz}^- $, $\mu_1 (\mathrm{k}_w)= J_{yw}^+ -J_{yw}^- $, $\mu_0$ is a chiral-breaking mass (in general can be $\tilde{\mathbf{k}}$ depended), and $\mathbf{\Gamma} = \{ \Gamma_0, \Gamma_1 ,\Gamma_2 ,\Gamma_3 , \Gamma_4  \}$ are five matrices,
\begin{align}	
	\Gamma_0 =\begin{pmatrix}     
		-\mathds{1} & 0 \\
		0 & \mathds{1} 
	\end{pmatrix}\,,
	\Gamma_1 =\begin{pmatrix}  
		0 &  \mathds{1} \\
		\mathds{1} & 0 
	\end{pmatrix}\,,
	\Gamma_2 =\begin{pmatrix}    
		0 & i\sigma_y\\
		-i\sigma_y & 0 
	\end{pmatrix}\,,
	\Gamma_3 =\begin{pmatrix}    
		0 &- i\sigma_z \\
		i\sigma_z & 0  
	\end{pmatrix}\,,
	\Gamma_4 =\begin{pmatrix}     
		0 & \frac{1}{2}\left(\sigma_x + i\sigma_y\right)\\
		\frac{1}{2}\left(\sigma_x - i\sigma_y\right) & 0   
	\end{pmatrix}\,, 
\label{eq:basis III}
\end{align}    
satisfying the chiral algebra, $\{\Gamma_0 , \Gamma_j\} = 2\delta_{0j}\mathds{1}$. 
As in Sec.~\ref{subsec:Model I}, the localised states around zero energy [cf.~Fig.~(3)(f) in the main text] are found by solving the continuoum Hamiltonian $\mathbf{{d}}_c\cdot\mathbf{\Gamma}$, where ${\textbf{d}}_c = \{\mu_0 , \mu_1 ,\epsilon \mu_2 , -{v} _x i\partial_x, -\epsilon{v}_y  i\partial_y \}$, with a plane wave ansatz $\psi = e^{i (\text{k}_x x + \text{k}_y y)}\phi$, where $\phi$ are the Bloch eigenvectors of ${\textbf{d}}_0 \cdot\bm{\Gamma}$. 
Placing our system on a semi-infinite space in $\mathbb{R}^2 _{\ge 0}$, where $x\ge0$ and $y\ge0$ (with Dirichlet boundary condition for $x<0$ and $y<0$) and expanding up to first order in $\mu_0$ we find a single normalisable solution below zero energy (assuming $\frac{\mu_1}{v_x} > 0$ and $\frac{\mu_2}{v_y} >0$)
\begin{align}
	\psi =e^{-\int\limits_0 ^x \frac{m_1}{v^+ _x} \text{d}x'}e^{-\int\limits_0 ^y \frac{m_2}{v^\pm _y} \text{d}y'}\begin{pmatrix}     
		1 \\
		0 \\
		0 \\
		0
	\end{pmatrix}\,,
	\label{eq:localised solution III}
\end{align}
with negative eigenvalue $- \mu_0$.
The above wavefunction is localised in both dimensions around the origin of the semi-infinite space and decays exponentially in space.

Following te same procedure as in Sec.~\ref{subsec:Model I}, we construct the interface between the macroscopic material described by $\hat{h}^{0}_{xy}(\tilde{\mathbf{k}})$ and the vacuum by extending the semi-infinite space $\mathbb{R}^2 _{\ge 0}$ to $\mathbb{R}^2$ and introducing domain walls in the $\mu_1 (x)$ and $\mu_2 (y)$ parameters.
The corresponding (2+1)D Lagrangian is given by
\begin{align}
	\mathcal{L} =\bar{\psi} \left(i\slashed{\partial} - \bm{\mu}\cdot \bm{\tau}\right)\psi \,,
\end{align}
where we adopt the same notation as in Sec.~\ref{subsec:Model I}.
We calculate Eq.~\eqref{eq:charge accumulation} up to second order in the gradients of $\mu_1$ and $\mu_2$ to find that the first nonvanishing contribution to the charge accumulation in a closed region $C$ [e.g., see Fig.~3(a)] is given by
\begin{align}
	q_C =\frac{i \epsilon^2}{2\pi^3}\int_{\tilde{C}}\int
	\frac{\mu_0\partial_{x} \mu_1 \partial_{y} \mu_2}{\left(-\omega^2 + \text{k}_x ^2 + \mu_1 ^2 + \mu_0\right)^3} \,\mathrm{d}\omega
	\mathrm{d}^2 \mathbf{{k}}\mathrm{d}^2 \mathbf{r}\,,
	\label{eq:corner charge III}
\end{align}
where we have expanded around $\epsilon\sim 0$ and kept up to second order. 
Changing variables of integration in Eq.~\eqref{eq:corner charge III}, $\mathbf{r}\to \mathbf{k}$, we obtain the expression of the \Snd Chern flux attached to a region in the 4D BZ of the ancestor model [cf., Eq.~\eqref{eq:2nd CN vector}].
In the limit $\mu_0\to 0$, i.e., when chiral symmetry is restored, the charge accumulation becomes quantized to
\begin{align}
	\lim_{\mu_0\to0} |q_C | =&
	\frac{1}{2} \text{   or   }		0\,.
\end{align}
In the nontrivial case $|q_C| = \frac{1}{2}$, the integration region defined by $\tilde{C}$ encloses a 4D crossing point in the $(\mathrm{k}_x , \mathrm{k}_y , \mathrm{k}_z ,\mathrm{k}_w)$-parameter space.

\section{Multipole description of macroscopic materials}
In this Section we define a macroscopic 2D material and for completeness show the derivation of the charge density following the modern approach of electric multipole expansion~\cite{king1993theory,vanderbilt1993electric,benalcazar2017quantized}. 
As an example, we consider the 2D pump family $\hat{h}^{\pi}_{xy}(\tilde{\mathbf{k}})$ and derive the charge transport under the adiabatic evolution of $\tilde{\mathbf{k}}$.
This provides the connection between the \Snd Chern number $c_2$ and the change of quadrupole moment $\partial_{\mathrm{k}_z} \partial_{\mathrm{k}_w} Q_{xy}(\tilde{\mathbf{k}}) $.

\subsection{Multipole expansion}
We assume that the material can be broken into unit cells $v(\mathbf{r})$ at positions $\mathbf{r}$, in which we define the multipole densities
\begin{align}
	\rho(\mathbf{r}) =& \frac{1}{v(\mathbf{r})}\int \limits_{v(\mathbf{r})} \text{d}^2 \mathbf{R} \rho(\mathbf{r}+\mathbf{R}) \\
	p_i(\mathbf{r}) = &\frac{1}{v(\mathbf{r})}\int \limits_{v(\mathbf{r})} \text{d}^2 \mathbf{R} \rho(\mathbf{r}+\mathbf{R})R_i \\
	q_{ij}(\mathbf{r}) =& \frac{1}{v(\mathbf{r})}\int \limits_{v(\mathbf{r})} \text{d}^2 \mathbf{R} \rho(\mathbf{r}+\mathbf{R})R_i R_j \,,
	\label{eq:densities}
\end{align}
where $\rho(\mathbf{r})$, $p_i (\mathbf{r})$ and $q_{ij}(\mathbf{r})$ are the charge, dipole and quadrupole densities over the unit cell at position $\mathbf{r}$ and the integrals run over $v(\mathbf{r})$. 
The electric potential at position $\mathbf{r}$ due to a charge distribution is given by
\begin{align}
\phi(\mathbf{r}) = \frac{1}{4\pi\epsilon_0}\sum\limits_{\mathbf{r}\,'} \int\limits_{v(\mathbf{r}\,')} \text{d}^2 \mathbf{R} \frac{\rho(\mathbf{r}\,'+\mathbf{R})}{|\mathbf{r}-\mathbf{R}-\mathbf{r}\,'|}
\end{align}
where $\epsilon_0$ is the dielectric constant, $\mathbf{r}\,'$ labels the unit cells and the integral runs over $v(\mathbf{r})$.
In the limit where the unit cell is small compared to the size of the material, $\mathbf{r}$ is treated as a continuous parameter and the sum over the positions of the unit cells becomes an integral over the area $V$ of the material $\sum_{\mathbf{r}'}\to \int_{V} \text{d}^{2}\mathbf{r}'$.
In this case, the expression for the electric potential can be expanded in powers of $1/|\mathbf{r}-\mathbf{r}\,'|$  
\begin{align}
\phi(\mathbf{r}) = \sum\limits_{l = 0}^{\infty} \phi^{l} (\mathbf{r})\,,
\end{align}
where
\begin{align}
\phi^{l} (\mathbf{r}) = \frac{1}{4\pi\epsilon}\int_{V} \text{d}^{2}\mathbf{r}\,' \int\limits_{v(\mathbf{r}\,')} \text{d}^2 \mathbf{R}\rho(\mathbf{r}\,'+\mathbf{R})\frac{|\mathbf{R}|^l}{|\mathbf{r}-\mathbf{r}\,'|^{l+1}}P_l \left(\frac{\mathbf{r}-\mathbf{r}\,'}{|\mathbf{r}-\mathbf{r}\,'|}\cdot \frac{\mathbf{R}\,'}{|\mathbf{R}\,'|}\right)\,,
\end{align}
and $P_l (x)$ is the $l$'th Legendre polynomial. 
Using the definitions of the multipole densities (see Eq.~\eqref{eq:densities}), the contributions to the total potential up to the quadrupole moment are given by~\cite{benalcazar2017quantized}
\begin{align}
\phi^{0} (\mathbf{r}) =& \frac{1}{4\pi\epsilon_0}\int_{V} \text{d}^{2}\mathbf{r}' 
 																			\rho(\mathbf{r}')\frac{1}{|\mathbf{r}-\mathbf{r}'|}\,,\\
\phi^{1} (\mathbf{r}) =& \frac{1}{4\pi\epsilon_0}\int_{V} \text{d}^{2}\mathbf{r}'
																			\left(-\partial_i p_i(\mathbf{r}')\frac{1}{|\mathbf{r}-\mathbf{r}'|}\right)
+ \frac{1}{4\pi\epsilon_0}\int_{\partial V} \text{d}^{1}\mathbf{r}'
																			\left(\hat{n}_i p_i(\mathbf{r}')\frac{1}{|\mathbf{r}-\mathbf{r}'|}\right)\,,\\
\phi^{2} (\mathbf{r}) = &\frac{1}{4\pi\epsilon_0}\int_{V} \text{d}^{2}\mathbf{r}'
																\left(\frac{1}{2}\partial_i \partial_j q_{ij}(\mathbf{r}')\frac{1}{|\mathbf{r}-\mathbf{r}'|}\right)
+ \frac{1}{4\pi\epsilon_0}\int_{\partial V} \text{d}^{1}\mathbf{r}'
																\left(-\hat{n}_i \partial_j q_{ij}(\mathbf{r}')\frac{1}{|\mathbf{r}-\mathbf{r}'|}\right)
+ \frac{1}{4\pi\epsilon_0}\sum_{p}
																\left(\frac{1}{2}\hat{n}^{\alpha}_i \hat{n}^{\beta}_j (1-\delta_{ij}) q_{ij}(\mathbf{r}')\frac{1}{|\mathbf{r}-\mathbf{r}'|}\right)\,,
\end{align}
where $\partial V$ is the 1D boundary (i.e., the edge) of the material and $p$ denotes the set of points where two edges, $\alpha$ and $\beta$, of $V$ intersect (i.e., the corners). 
The zeroth term in the expansion $\phi^{0} (\mathbf{r})$ is proportional to the bulk charge density of the material, the first-order term $\phi^{1} (\mathbf{r})$ is proportional to (i) the areal charge density created by the divergence of a bulk dipole moment and (ii) the line charge density $\sigma_{\alpha} = \hat{n}^{\alpha}_i p_i(\mathbf{r})$ created by a free dipole moment on edge $\alpha$ with normal vector $\hat{\vec{n}}^{\alpha}$, the second-order term $\phi^{2} (\mathbf{r})$ is proportional to (i) the areal charge density created by a bulk quadrupole moment, (ii) the line charge density $\lambda_{a} = -\hat{n}^{a}_i \partial_j  q_{ij}$ created by the divergence of the quadrupole moment on an edge $\alpha$ with normal vector $\hat{\vec{n}}^{\alpha}$ and (iii) the free point charges $\eta_{\alpha,\beta} =\frac{1}{2}\hat{n}^{\alpha}_i \hat{n}^{\beta}_j Q_{ij}$ at the corners of the material where two edges, $\alpha$ and $\beta$, intersect. 

The total charge density at a position $\mathbf{r}$ due to an electric potential $\phi(\mathbf{r})$ is given by 
\begin{align}
\rho_{0}(\mathbf{r}) = -\nabla^2 \phi(\mathbf{r})\,.
\end{align}%
Using the identity $\nabla^2 \left(\frac{1}{|\mathbf{r}-\vec{r'}|}\right) = -4\pi\delta^{(2)}(\mathbf{r}-\vec{r'})$, the total charge density can be expressed as $\rho_{0}(\mathbf{r}) = \rho_{\text{bulk}}(\mathbf{r})+\rho_{\text{edge}}(\mathbf{r})+\rho_{\text{corner}}(\mathbf{r})$ where
\begin{align}
\rho_{\text{bulk}}(\mathbf{r}) =& \rho(\mathbf{r})  - \partial_i p_i(\mathbf{r}) + \frac{1}{2}\partial_i \partial_j q_{ij}(\mathbf{r})\,,\\
\rho_{\text{edge}}(\mathbf{r}) =& \sigma_{\alpha}  + \lambda_{a}\,,\\
\rho_{\text{corner}}(\mathbf{r}) =& \eta_{\alpha,\beta}\,.\\
\end{align}%

\subsection{Quadrupole winding in a 2D topological pump}
We consider the 2D pump family $\hat{h}^{\pi}_{xy}(\tilde{\mathbf{k}})$ with an on-side staggered potential with strength $\mu_0$. 
The only nonzero multipole of this model is the bulk quadrupole $Q_{xy}(\tilde{\mathbf{k}})$. 
We assume $\partial_t\mathrm{k}_z = E$ and $\partial_x\mathrm{k}_w = B$, where $E$ and $B$ are small parameters.
In other words, $\mathrm{k}_z $ is adiabatically evolved as a function of time $t$ with period $1/E$ and $\mathrm{k}_w $ is smoothly changing over the $x$-direction with period $1/B$. 
Under this assumption, the resulting charge transport after a period across the $y$-direction of the 2D material is given by
\begin{eqnarray}
\Delta q = \int\mathrm{d}t\int\mathrm{d}x\partial_t \partial_x Q_{xy}.
\end{eqnarray}
Changing variables of integration $ t\to \mathrm{k}_z$ and $ x\to \mathrm{k}_w$ we find that the total charge transport after a period is given by
\begin{eqnarray}
\Delta q = \int\mathrm{d}\mathrm{k}_z\int\mathrm{d}\mathrm{k}_w\partial_{\mathrm{k}_z} \partial_{\mathrm{k}_w} Q_{xy}.
\end{eqnarray}

\section{The \Snd Chern number \label{apsemic}}

In this Section, we calculate the \Snd Chern flux of the three models defined in the main text and show its connection to the charge accumulation at the corner of the 2D descendant families (cf., Sec~\ref{sec:charge accumulation}). 
In general, the momentum-space representation of a chiral Hamiltonian is given by
\begin{eqnarray}
 \mathbf{{d}}\cdot\mathbf{\Gamma}= \begin{pmatrix}
-\mathrm{d}_0 \mathds{1} & M \\
M^\dagger & \mathrm{d}_0 \mathds{1}
\end{pmatrix}\,,
\end{eqnarray}
where $\textbf{{d}} = \{\text{d}_0 , \text{d}_1 ,\text{d}_2 ,\text{d}_3 ,\text{d}_4\}$ is a real-valued vector, $M$ is a 2x2 matrix and $\mathbf{\Gamma} = \{ \Gamma_0, \Gamma_1 ,\Gamma_2 ,\Gamma_3 , \Gamma_4 \}$ are five matrices satisfying the chiral algebra, $\{\Gamma_i , \Gamma_0\} = 2\delta_{i0}\mathds{1}$. 
The corresponding energy bands and Bloch vectors are given by 
\begin{align}
	E^\pm _i = \pm \sqrt{\mathrm{d}_0 ^2 + \mathcal{E}_i}
\end{align}
and
\begin{align}
	U^\pm _i =\begin{pmatrix}     
		\frac{M u_i}{E^\pm _i + \text{d}_0 } \\
		u_i
	\end{pmatrix}\,
\end{align}
where $\mathcal{E}_i$ and $u_i$ are the eigenvalues and eigenvectors of $M^\dagger M$.

The \Snd Chern flux over a specified region $\Omega$ is defined as the energy-momentum integral of the symmetric form 
\begin{eqnarray}
\Phi_2 = -\frac{\pi^2}{15}\epsilon_{\mu\nu\rho\sigma\tau}\int_{\Omega}\frac{\text{d$^4$k}\text{d}\omega}{\left(2\pi\right)^5}\,\text{Tr}\,\left(G\partial_{q_\mu} G^{-1}\right)\left(G\partial_{q_\nu} G^{-1}\right)\left(G\partial_{q_\rho} G^{-1}\right)\left(G\partial_{q_\sigma} G^{-1}\right)\left(G\partial_{q_\tau} G^{-1}\right)
\label{eq:2nd CN}
\end{eqnarray}
where $q_\mu = \{\omega,\text{k}_x , \text{k}_y , \text{k}_z , \text{k}_w\}$ and $G = \left[\omega + i \delta -  \mathbf{d}\cdot\mathbf{\Gamma}\right]^{-1}$ is the single-particle Green's function. 
We consider 4D Hamiltonians where $\mathbf{d}=\{\mu_0 (\tilde{\mathbf{k}}), \mu_1 (\tilde{\mathbf{k}}), \mu_2 (\tilde{\mathbf{k}}), {v}_x (\tilde{\mathbf{k}}) \mathrm{k}_x, {v}_y (\tilde{\mathbf{k}}) \mathrm{k}_y  \}$, and, for simplicity, take $\mu_0$, $v_x$, and $v_y$ to be constants, $\mu_1 (\tilde{\mathbf{k}})\equiv\mu_1 (\text{k}_z)$ to be a function of only $\text{k}_z$ and $\mu_2 (\tilde{\mathbf{k}})\equiv\mu_2 (\text{k}_w)$ to be a function of only $\text{k}_w$. 

\subsection{Model I}
We consider the 4D Hamiltonian given in Eq.~\eqref{eq:Model I}, where $\mathbf{\Gamma} = \{ \Gamma_0, \Gamma_1 ,\Gamma_2 ,\Gamma_3 , \Gamma_4 \}$ are the five matrices given in Eq.~\eqref{eq:basis I}. 
In this case, Eq.~\eqref{eq:2nd CN} is given by
\begin{eqnarray}
\Phi_2 = -\frac{i}{\pi^3}\int_{\Omega} \frac{\mu_0 \partial_{\text{k}_z}\mu_1 \partial_{\text{k}_w}\mu_2}{(\omega^2 - |\mathbf{{d}}|^2)^3}\,\mathrm{d}^2 \mathbf{{k}}\mathrm{d}^2 \tilde{\mathbf{{k}}} \text{d}\omega\,.
\end{eqnarray}
Integrating over $\omega$ using the residue theorem we obtain
\begin{eqnarray}
\Phi_2 = \frac{3}{8\pi^2}\int\frac{\mu_0 \partial_{\text{k}_z}\mu_1 \partial_{\text{k}_w}\mu_2}{|\mathbf{{d}}|^{5/2}}\,\mathrm{d}^2 \mathbf{{k}}\mathrm{d}^2 \tilde{\mathbf{k}}\,.
\end{eqnarray}
Generalizing this to a Hamiltonian $\mathbf{{d}}\cdot\mathbf{\Gamma}$, where $\mathbf{{d}}$ is now a smooth function of ${\mathbf{k}}$ and $\tilde{\mathbf{k}}$, we obtain
\begin{eqnarray}
\Phi_2 =\frac{3}{8\pi^2}\int\hat{\mathbf{\textbf{d}}}\cdot (\partial_{\mathrm{k}_x}\hat{\textbf{\text{d}}} \times \partial_{\mathrm{k}_y}\hat{\mathbf{\textbf{d}}}\times \partial_{\mathrm{k}_z}\hat{\mathbf{\textbf{d}}} \times \partial_{\mathrm{k}_w}\hat{\mathbf{\textbf{d}}})\mathrm{d}^2 \mathbf{{k}}\mathrm{d}^2 \tilde{\mathbf{k}}\,.
\label{eq:2nd Chern Flux}
\end{eqnarray}
The above expression is equal to the charge accumulation $q_C$ derived in Eq.~\eqref{eq:corner is 2nd CN}. 
Importantly, in the limit where the chiral mass $\mu_0$ goes to zero, the \Snd Chern flux $\Phi_2$ takes quantised values 
\begin{align}
 \lim_{\mu_0 \to 0^\pm}|\Phi_2|=\frac{1}{2}\text{  or  }0\,,
\end{align}
depending on if the integration domain $\Omega$ encloses the singular point in the spectrum where the gap size becomes proportional to $\sim\mu_0$.

\subsection{Model II}

We consider the 4D Hamiltonian given in Eq.~\eqref{eq:Model II}, where $\mathbf{\Gamma} = \{ \Gamma_0, \Gamma_1 ,\Gamma_2 ,\Gamma_3 , \Gamma_4 \}$ are the five matrices given in Eq.~\eqref{eq:basis II}. 
Taking the limit $\epsilon\to 0$, Eq.~\eqref{eq:2nd CN} is given by
\begin{eqnarray}
\lim\limits_{\epsilon\to 0}\Phi_2 =\frac{i \epsilon^2}{3\pi^3}\int_{\Omega}
\frac{\omega^2 +\text{k}_x ^2 + \mu_1 ^2 - \mu_0 ^2}{\left(-\omega^2 +\text{k}_x ^2 + \mu_1 ^2 + \mu_0\right)^4} \mu_0\partial_{x} \mu_1 \partial_{y} \mu_2\,\mathrm{d}^2 \mathbf{{k}}\mathrm{d}^2 \tilde{\mathbf{{k}}} \text{d}\omega\,.
\label{eq:2nd Chern Flux II}
\end{eqnarray}
The above expression is proportional to the charge accumulation $q_C$ derived in Eq.~\eqref{eq:corner charge II}. 
%Importantly, in the limit where the chiral mass $\mu_0$ goes to zero, the \Snd Chern flux $\Phi_2$ takes quantised values 
%
%\begin{align}
%	\lim_{\mu_0 \to 0^\pm}|\Phi_2|=\frac{1}{2}\text{  or  }0\,,
%\end{align}
%depending on if the integration domain $\Omega$ encloses the singular point in the spectrum where the gap size becomes proportional to $|\mathbf{{d}}|\sim\mu_0$.

\subsection{Model III}

We consider the 4D Hamiltonian given in Eq.~\eqref{eq:Model III}, where $\mathbf{\Gamma} = \{ \Gamma_0, \Gamma_1 ,\Gamma_2 ,\Gamma_3 , \Gamma_4 \}$ are the five matrices given in Eq.~\eqref{eq:basis III}. 
Taking the limit $\epsilon\to 0$, Eq.~\eqref{eq:2nd CN} is given by
\begin{eqnarray}
\lim\limits_{\epsilon\to 0}\Phi_2 =\frac{i \epsilon^2}{2\pi^3}\int_{\Omega}
\frac{\mu_0\partial_{x} \mu_1 \partial_{y} \mu_2}{\left(-\omega^2 + \text{k}_x ^2 + \mu_1 ^2 + \mu_0^2\right)^3}\,\mathrm{d}^2 \mathbf{{k}}\mathrm{d}^2 \tilde{\mathbf{{k}}} \text{d}\omega\,.
\label{eq:2nd CN vector}
\end{eqnarray}
The above expression is equal to the charge accumulation $q_C$ derived in Eq.~\eqref{eq:corner charge III}. 
%Importantly, in the limit where the chiral mass $\mu_0$ goes to zero, the \Snd Chern flux $\Phi_2$ takes quantised values 
%
%\begin{align}
%	\lim_{\mu_0 \to 0^\pm}|\Phi_2|=\frac{1}{2}\text{  or  }0\,,
%\end{align}
%depending on if the integration domain $\Omega$ encloses the singular point in the spectrum where the gap size becomes proportional to $\sim\mu_0$.

\end{document}